\UseRawInputEncoding
\documentclass[usenatbib,useAMS,usedcolumn]{mnras}

\usepackage{epsfig}
\usepackage{times,pdflscape}
\pdfoutput=1
\hypersetup{pdfauthor={H.L. Liu},
            bookmarksnumbered=true}

\newcommand{\cm}{\ensuremath{\mbox{~cm}}}
\newcommand{\pcmsq}{\ensuremath{\cm^{-2}}}
\newcommand{\pcmcu}{\ensuremath{\cm^{-3}}}

\newcommand{\msun}{$M_{\odot}$}
\newcommand{\lsun}{$L_{\odot}$}
\newcommand{\um}{$\mu$m}

\newcommand{\egcite}{\citep[e.g.,][]}

\newcommand{\hcop}{HCO$^{+}$}
\newcommand{\htcop}{H$^{13}$CO$^{+}$}
\newcommand{\halpha}{H$40{\alpha}$}

\newcommand{\hii}{H\textsc{ii}}

\graphicspath{{./0figs/}}

\begin{document}

\title[High-mass star formation in HFS cloud
] 
{Evidence of high-mass star formation through multi-scale mass accretion in hub-filament-system clouds
}


\author[H.-L. Liu et al.]{
Hong-Li Liu,$^{\star 1}$
Anandmayee Tej,$^{\star 2}$
Tie Liu,$^{\star 3,4}$
Patricio Sanhueza,$^{5,6}$
Shengli Qin,$^{1}$
Jinhua He,$^{7,8,9}$
\newauthor
Paul F. Goldsmith,$^{10}$
Guido Garay,$^{9}$
Sirong Pan,$^{1}$
Kaho Morii,$^{5,11}$
Shanghuo Li,$^{12}$,
Amelia Stutz,$^{13,14}$
\newauthor
Ken'ichi Tatematsu,$^{5}$
Feng-Wei Xu,$^{15,16}$
Leonardo Bronfman,$^{9}$
Anindya Saha,$^{2}$
Namitha Issac,$^{17}$
Tapas Baug,$^{18}$
\newauthor
L. Viktor Toth,$^{19}$
Lokesh Dewangan,$^{20}$
Ke Wang,$^{15,16}$
Jianwen Zhou,$^{21}$
Chang Won Lee,$^{22,23}$
Dongting Yang,$^{1}$
\newauthor
Anxu Luo,$^{1}$
Xianjin Shen,$^{1}$
Yong Zhang,$^{24}$
Yue-Fang Wu,$^{15,16}$
Zhiyuan Ren,$^{21}$
Xun-Chuan Liu,$^{3}$
\newauthor
Archana Soam,$^{25}$
Siju Zhang,$^{15,16}$
Qiu-Yi Luo,$^{3}$
\\
Affiliations are listed at the end of the paper}

\date{Accepted 2023 January 3; Received 2022 December 6; in original form 2022 September 13}

\pagerange{\pageref{firstpage}--\pageref{lastpage}} \pubyear{2021}

\maketitle

\label{firstpage}

\begin{abstract} 
We present a statistical study of a sample of 17 
hub-filament-system (HFS) clouds of high-mass star formation
using high-angular resolution ($\sim 1$--2\arcsec) ALMA 1.3\,mm and 3\,mm continuum data.
The sample includes 8 infrared (IR)-dark and 9 IR-bright types, which correspond to an evolutionary sequence from the IR-dark to IR-bright stage.
The central massive clumps and their associated most massive cores are observed to follow a trend of increasing
mass ($M$) and mass surface density ($\Sigma$) with evolution from IR-dark to IR-bright stage.
In addition, a mass-segregated cluster of young stellar objects (YSOs) are revealed in both IR-dark and IR-bright HFSs with 
 massive YSOs located in the hub and the  population of low-mass YSOs distributed over larger areas.
Moreover, 
outflow feedback in all HFSs are found to escape preferentially through the inter-filamentary diffuse cavities, suggesting that outflows would
render a limited effect on the disruption of the HFSs and ongoing high-mass star formation therein. 
From the above observations, we suggest that high-mass star formation in the HFSs can be described by a multi-scale mass accretion/transfer scenario, from hub-composing filaments
through clumps down to cores, that can naturally lead to a mass-segregated cluster of stars.

\end{abstract} 


\footnotetext[1]{E-mail: hongliliu2012@gmail.com; tej@iist.ac.in; liutie@shao.ac.cn}

\section{Introduction} \label{sec:intro}
High-mass stars are fundamental components of
the universe, significantly impacting a multitude of 
 astrophysical processes, for instance,  the structure and evolution of the 
universe and its constituents \egcite{Lar03}. 
Understanding how high-mass stars form has therefore long been an active area of astrophysical research \egcite{Mot18}. 
It is generally accepted that high-mass stars form in clusters probably as a result of the hierarchical, multi-scale fragmentation process 
from clouds, through filaments, clumps and cores, down to seeds of star formation 
\egcite{Zha09,Wan11,Wan14,Beu18,Yua18,Vaz19,Li 19, Li 20b,Pad20,Kum20,Liu22a,Liu22b,Hac22,Che22}.
Under this paradigm, filaments as a ``conveyor belt'' play a critical role in transporting gas material between the scales above and below
\egcite{Lon14,Vaz19,Pad20,Kum20}.

Indeed, filaments have been observed to be ubiquitous in the interstellar medium (ISM), and the sites of star formation in both nearby low-mass and distant 
high-mass star-forming clouds \egcite{And10,And19,Mol10,Stu16,Guo21,Yua21}.
Crisscrossing filaments result in a special web that  can comprise of three or more filaments converging toward a central web node. Defined as hub-filament systems (HFSs), these web networks are  considered as a unique category of filaments for star formation, especially for high-mass star formation \egcite{Mye09,Kum20}.
In the definition of HFS, the web node is defined as the hub while the associated individual filaments {are defined} as the hub-composing filaments.
In general, the central hub has a lower aspect ratio but a higher column density, which are in contrast to the high aspect ratio and low column density observed in the hub-composing filaments \egcite{Mye09,Kum20}.

The hierarchical density structure discussed above can promote high-mass star formation. This is supported by several observational studies that include both single dish and interferometric observations from the far-infrared (IR) to  millimeter regime
\egcite{Sch12,Liu12,Per13,Wil18,Yua18,Iss19,Kum20,And21,Bel22,Liu22a,Liu22b,San21,Sah22,Zho22,Tho22}. These studies reveal that young massive stellar clusters appear in HFSs with the high-mass stars being preferentially formed in the hubs. 
Longitudinal gas flows along the hub-composing filaments, which are observed in several HFSs to converge toward the hub
at typical flow rates of $\sim 10^{-4}$--$10^{-3}$\,\msun~yr$^{-1}$, have been demonstrated to account for the required mass accretion \egcite{Yua18,Che19,Tre19,Liu22a}. 

The above scenario advocates for further detailed observational studies of HFSs from the perspective of high-mass star formation and for the development of theoretical models.
For instance, the latest-generation models 
such as ``global hierarchical collapse'' (GHC, \citealt{Vaz19}) and ``inertial-inflow'' (I2, \citealt{Pad20}),
which in terms of the accretion process could be complementary to the two proposed competing theories of ``turbulent core accretion'' \citep{McK03} and ``competitive accretion'' \citep{Bon01}. The merits of these latest-generation models can be
attributed to the envisioned multi-scale gas accretion, from clouds to the seeds of star formation, which was proposed in earlier times as ``clump-fed" accretion in simulations of \citet{Wan10}, but not fully developed to the cloud scales due to the computation limitation at that time.
In addition, the HFSs are often reproduced to be a common signature in the multi-scale accretion models
as the preferential system of cluster and high-mass star formation.
However, the major driver of multi-scale gas accretion (e.g., gravity, and/or turbulence) is predicted to be different in different models \egcite{Liu22b}. This is inferred from our previous ALMA observations, at $\sim2$\arcsec\ resolution, toward a well-studied, high-mass star-forming filamentary IRDC, G034.43+00.24 (e.g., \citealt{San10,Sak13,Fos14,Liu20}, G34 hereafter).  G34 can be regarded as an HFS with the hub located at the MM1 clump (see Fig.\,1 of \citealt{Liu22a,Liu22b}). Our investigations have revealed multi-scale accretion process from cloud down to the seeds of star formation and the scale-dependent nature of gas kinematics of the multi-scale, hierarchical density structures. 
Interpreting our results
in the framework of both GHC and I2 models, we conclude
that the scale-dependent combined effect of turbulence and gravity is essential to explain the multi-scale, dynamical accretion process responsible for
high-mass star formation in G34. 

In this paper, we aim to carry out  a statistical study  of a sample of 17 high-mass star forming HFSs 
using high-angular resolution ($\sim$1--2\,\arcsec) ALMA continuum data. The sample was selected particularly
 to contain two different infrared (IR) characteristics (i.e., 8 IR-dark and 9 IR-bright objects) since these two IR types can represent two different evolutionary stages (see Sect.\,\ref{sec:sample}). 
 The purpose of this study is to gain insights into high-mass star formation scenarios in HFS clouds by analysing the hierarchical structures of the HFSs as a function of evolution of high-mass star formation. 
The paper is organised as follows: Sect.\,2 briefly describes the selection of the HFS sample and the ALMA data used, 
Sect.\,3 presents analysis on the hierarchical structures of the HFSs (i.e., clouds, clumps, and cores), star formation therein,
and the effect of outflow feedback on star formation. Sect.\,4 discusses the multi-scale accretion scenario,  
from the core through clump up to the cloud scales,  and Sect.\,5 gives a comprehensive summary of the results obtained.

\section{Sample and ALMA data}
\label{sec:observe}

\begin{figure*}
\centering
\includegraphics[width=6.8 in]{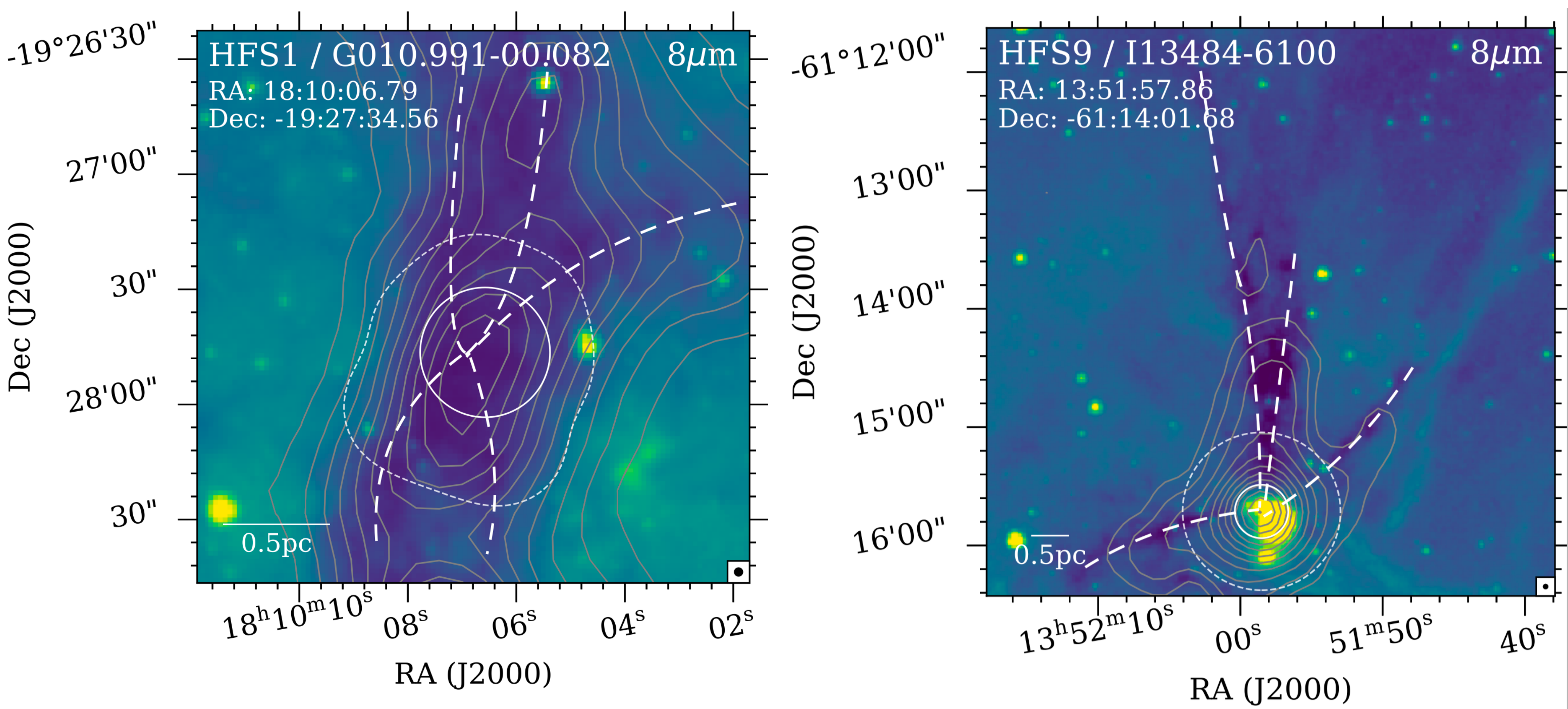}
\caption{  Images showing examples of the IR-dark (left) and bright (right) HFS clouds at Spitzer 8\,\um.
The  contours represent 870\,\um\ dust continuum  from the ATLASGAL survey \citep{Sch09}. 
The solid circles represent the compact dust clumps (Sect.\,\ref{sec:clumps}). 
The dashed loop/circle demarcates the central subcloud field covered by our ALMA observations. The dashed curves 
identify the filamentary structures.
The 8.0\,\um\ beam (i.e., 2\arcsec) are shown at the bottom right-hand corner of the corresponding panel.
The image size for each HFS is determined individually to recover the full view of its HFS morphology at 8\,\um. 
}
\label{fig:overview_hfs}
\end{figure*} 

\begin{table*}
\centering
\caption{Parameters of the HFSs and their central clumps.}
\label{tab:hfs_cloud}
\resizebox{17cm}{!}{
\begin{tabular}{clllllllll}
\hline\hline
           HFS cloud &                Alias &              RA &             DEC &             $d$ &   $L_{\rm bol}$ &  $T_{\rm dust}$ & $F_{\rm 870\mu m}^{int.}$ & $M_{\rm clump}$  & $\Sigma_{\rm clump}$   \\
              ID &                      &  J2000 &  J2000 &    kpc & log($L_{\rm \odot}$) &      K &     Jy & $M_{\odot}$ & g cm$^{-2}$    \\

\hline
   1  &   G010.991-00.082  &       18:10:06.58  &      -19:27:46.44  &     3.7  &     1.7  &    12.0  &    1.94  &     355  &    0.38  \\ 
   2  &   G014.492-00.139  &       18:17:22.08  &      -16:24:59.40  &     3.9  &     2.5  &    13.0  &    3.89  &     685  &    0.73  \\ 
   3  &   G028.273-00.167  &       18:43:31.27  &       -4:13:18.48  &     5.1  &     1.6  &    12.0  &    1.05  &     365  &    0.39  \\ 
   4  &   G331.372-00.116  &       16:11:34.08  &      -51:34:56.28  &     5.4  &     1.6  &    14.0  &    0.90  &     266  &    0.28  \\ 
   5  &   G340.222-00.167  &       16:48:30.74  &      -45:11:04.56  &     4.0  &     1.6  &    15.0  &    1.07  &     155  &    0.17  \\ 
   6  &   G340.232-00.146  &       16:48:27.46  &      -45:09:48.24  &     3.9  &     1.1  &    14.0  &    1.16  &     179  &    0.19  \\ 
   7  &   G341.039-00.114  &       16:51:14.21  &      -44:31:24.24  &     3.6  &     1.8  &    14.3  &    1.21  &     153  &    0.16  \\ 
   8  &   G343.489-00.416  &       17:01:01.34  &      -42:48:07.92  &     2.9  &     0.8  &    10.3  &    1.04  &     156  &    0.17  \\ 
   9  &       I13484-6100  &       13:51:58.51  &      -61:15:42.84  &     5.4  &     4.8  &    31.8  &    4.10  &     600  &    0.64  \\ 
  10  &       I15394-5358  &       15:43:16.46  &      -54:07:15.24  &     1.8  &     3.7  &    34.0  &   29.69  &     482  &    0.51  \\ 
  11  &       I15520-5234  &       15:55:48.62  &      -52:43:05.88  &     2.6  &     5.1  &    32.2  &   38.82  &     733  &    0.78  \\ 
  12  &       I16272-4837  &       16:30:59.11  &      -48:43:53.40  &     2.9  &     4.3  &    23.1  &   17.64  &     748  &    0.80  \\ 
  13  &       I16351-4722  &       16:38:50.69  &      -47:27:58.68  &     3.0  &     4.9  &    30.4  &   25.43  &     649  &    0.69  \\ 
  14  &       I16424-4531  &       16:46:06.36  &      -45:36:44.64  &     2.6  &     3.9  &    24.6  &    6.70  &     153  &    0.16  \\ 
  15  &       I17016-4124  &       17:05:11.18  &      -41:29:05.28  &     1.4  &     5.3  &    32.0  &   51.16  &     975  &    1.04  \\ 
  16  &       I17233-3606  &       17:26:42.74  &      -36:09:18.00  &     1.3  &     4.6  &    29.9  &  123.33  &     201  &    0.21  \\ 
  17  &       I18507+0121  &       18:53:18.12  &       +1:25:24.24  &     3.7  &     3.5  &    22.7  &   13.34  &     882  &    0.94  \\ 

\hline
\end{tabular}
}
\begin{flushleft}
{\bf Note:} The distance $d$ of the HFSs  comes from \citet{San19} for HFSs\,1--8, and \citet{Liu21} for HFSs\,9--17. The bolometric luminosity $L_{\rm bol}$ of the HFSs
is approximately represented by that of their centrally located young stellar objects (YSOs, \citealt{Bro96,Con13,Liu20a,Liu20b,Liu21}). 
\end{flushleft}
\end{table*}

\subsection{Sample} \label{sec:sample} 
 The sample investigated here consists of 17 HFS clouds selected from the ASHES (The ALMA Survey of 70 $\mu$m Dark High-mass Clumps in Early Stages; \citealt{San19,Li 20, Sab22}) and the ATOMS (ALMA Three-millimeter Observations of Massive Star-forming regions; \citet{Liu20a,Liu20b,Liu21}) surveys. The selection is based on their morphological appearance in the Spitzer 8\,\um\ image (see Fig.\,\ref{fig:overview_hfs}). The selected HFS cloud is required to be globally seen as an HFS morphology in 8\,\um\ emission with at least three hub-composing filaments intersecting at the central hub (see Fig.\,\ref{fig:overview_hfs}). The hub-composing filaments appear as elongated dark lanes against bright 8\,\um\ background emission. With the matching angular resolution (i.e., 2\arcsec) as that of the ALMA data, the Spitzer image facilitates the ALMA analysis of the identified HFSs.

The selected HFSs are classified into 8 IR-dark and 9 IR-bright HFS clouds based on the lack or presence of IR emission in the central hubs. The IR-dark sample (i.e., HFSs\,1--8) is from the ASHES survey where the hubs are identified as IR-dark clumps from 3.6 to 70\,\um\ \citep{San19}. In contrast, the hubs of the IR-bright HFSs (i.e., HFSs\,9--17), selected from the ATOMS survey, are bright in the same IR regime. This, in essence, is a reflection of the bolometric luminosity ($L_{\rm bol}$) of the embedded young stellar objects (YSOs) in the hubs.  Making a simple assumption, the $L_{\rm bol}$ of YSOs in these hubs is assumed to 
approximately represent that of their host HFSs (see Table\,\ref{tab:hfs_cloud}).
While the absence of compact IR emission can qualify the central hubs of IR-dark HFSs as prestellar candidates, 
being 70\,\um\ dark does not always imply absolute lack of star formation \egcite{Li 19, Li 20, Mor21, Taf21, Sak22}. Hence, the lack of YSOs here could also indicate a relatively quiescent star formation stage with very low bolometric luminosities (i.e., $L_{\rm bol}\la300$\,\lsun, see Table\,\ref{tab:hfs_cloud}).
On the other hand, the central hubs of the IR-bright
HFSs have high-luminosity ($L_{\rm bol}\ga10^4$\,\lsun, see Table\,\ref{tab:hfs_cloud}) IRAS sources typical of high-mass stars, thus suggesting an active star formation stage. 
 Interestingly, even with the relatively limited sample of HFSs investigated in this study strikingly different regimes of $L_{\rm bol}$/$M_{\rm clump}$ ratios are seen for the IR-dark and IR-bright HFSs (see Sect.\,\ref{sec:clumps}). This lends strong support to the inference that two different star-formation stages are probed with these two IR types.

The basic parameters (e.g., source name, distance, and luminosity) of the selected sample are listed in Table\,\ref{tab:hfs_cloud}. 
They have a median distance of $\sim3.6$\,kpc in a range of $\sim 1.3$--5.4\,kpc with the IR-dark HFSs ($\sim3.9$\,kpc) being on average around 1.4 times farther than 
the IR-bright ones ($\sim 2.7$\,kpc). It is worth noting that the 9 IR-bright HFSs selected here have been reported by \citet{Zho22} as part of a statistical study to identify and study
HFSs in a large sample of 146 active massive protoclusters 
based on the \htcop~(1--0) line data from the same ATOMS survey.
Among the sample studied by \citet{Zho22}, these 9 IR-bright HFSs stand out in terms of the  8\,\um\ appearance of their hub-composing filaments as elongated dark lanes.

Figure\,\ref{fig:overview_hfs} presents an example of the large-scale appearance of the selected 17 HFSs studied here in the 
Spitzer 8\,\um\ image.  Overlaid in  gray contours on the image is ATLASGAL 
870\,\um\ continuum representative of cold, dense dust thermal emission. The size of the images
was adjusted to recover the complete presence of the global HFS appearance in 8\,\um\ emission.
As shown in the figure, the global HFS appearance can be identified for all 17 HFS clouds with the hub-composing 
filaments  intersecting at the centrally located hub.
The hub-composing filaments seen as elongated dark lanes at 8\,\um\  have associated
870\,\um\ dust emission representative of cold and dense material, indicating that the filaments are essentially high density structures.

\subsection{ALMA data}
\label{sec:almadata}
We made use of the combined 12m+7m continuum data from  ASHES (project IDs: 2015.1.01539.S, PI: Patricio Sanhueza), and ATOMS (project ID: 2019.1.00685.S; PI: Tie Liu) surveys. Detailed discussion on the scientific goals, observing set-ups, and data reduction can be found in \citet{San19,Liu20a,Liu20b,Liu21}. 
Briefly, the two surveys observed the central area of radius $\sim31$--$39$\arcsec\ of our selected HFSs (see the dashed loop in Fig.\,1)
in different observing modes at different wavelengths, i.e., 
the 1.3\,mm mosaic mode for ASHES, and 3\,mm pointing observing mode for ATOMS. In addition,
the synthesized beams of the combined 12m+7m continuum data of the two surveys are different
with $\sim1.2$\arcsec\ for ASHES and $\sim2.0$\arcsec\ for ATOMS.
However, the two surveys have similar field of views (FoVs, i.e., $\sim62$\arcsec\ for ASHES and $78$\arcsec\ for ATOMS).
Given their typical distances (see Sect.\,\ref{sec:sample}), 
the IR-dark and bright HFSs selected from the respective ASHES, and ATOMS surveys have very close 
linear-scale FoVs (i.e., $\sim 1.0$\,pc and $\sim 1.2$\,pc, respectively). This rather close  agreement of FoVs ensures detailed high-resolution analysis (e.g., for cores)
over almost the same spatial scales for most of the HFSs studied here.
The ASHES and ATOMS surveys have a maximum angular recoverable scale of $19$\arcsec\ and 60\arcsec, respectively, for the combined data.
Note that the different angular resolutions and maximum recoverable scales may lead to some observation biases to the properties of cores (e.g., mass),
which will be discussed in Sect.\,\ref{sec:core_params_derived}.
In addition, the typical sensitivities of the combined continuum data for ASHES and ATOMS 
are $\sim0.1$\,mJy beam$^{-1}$ \citep{San19} at 1.3\,mm and $\sim0.3$\,mJy beam$^{-1}$ at 3\,mm \citep{Liu22a}, respectively,
which correspond to a mass sensitivity of $\sim0.04$\,\msun\ (for a temperature of 15\,K typical of IR-dark cases), and $\sim1.2$\,\msun\ (for 25\,K typical of IR-bright cases), 
respectively, at a typical distance of 3.6\,kpc (see Sect.\,\ref{sec:clumps} for mass calculation).

\section{Results and analysis}
\label{sec:HFS_analy}

\begin{figure*}
\centering
\includegraphics[width=2.35 in]{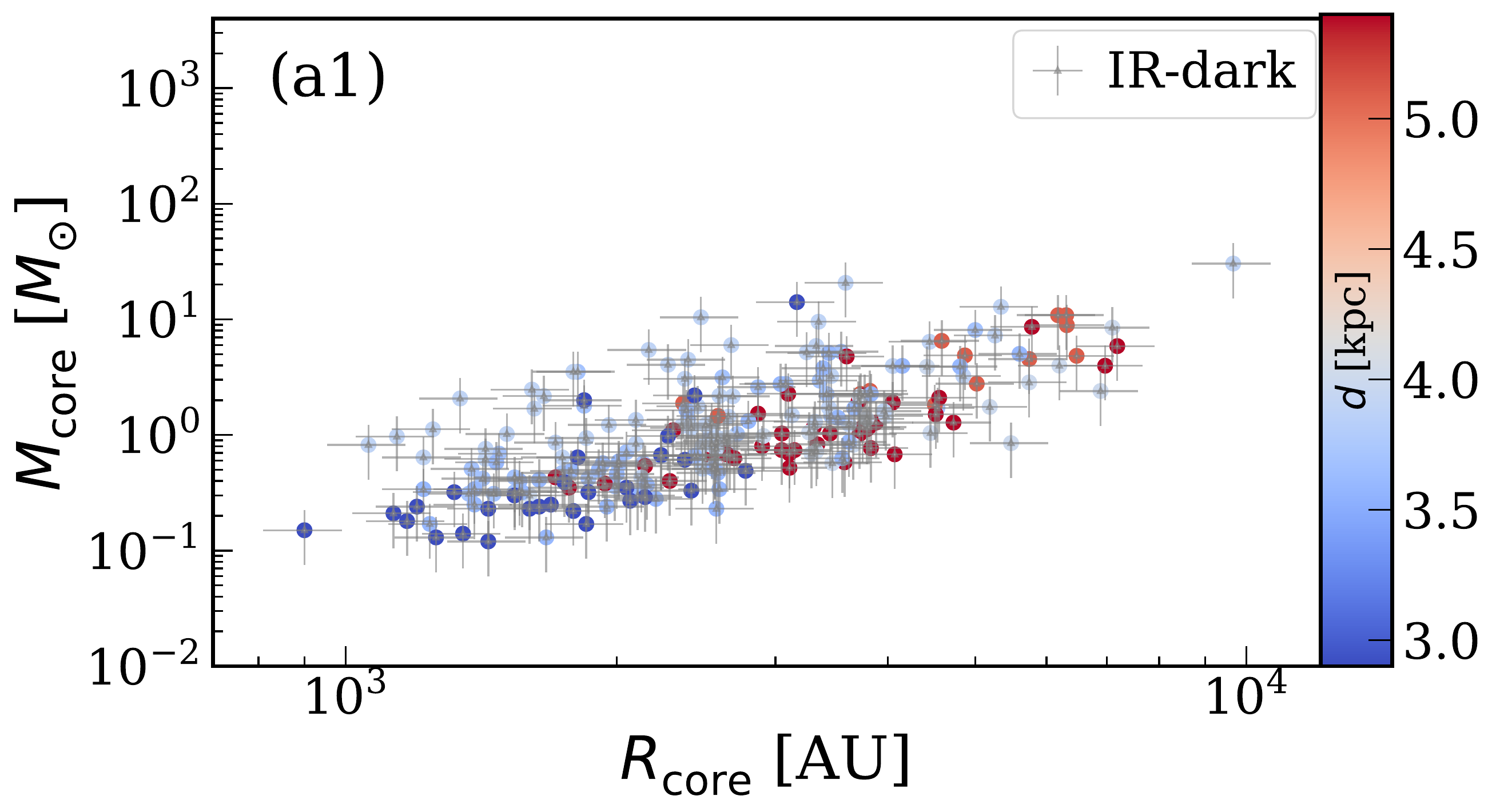}
\includegraphics[width=2.35 in]{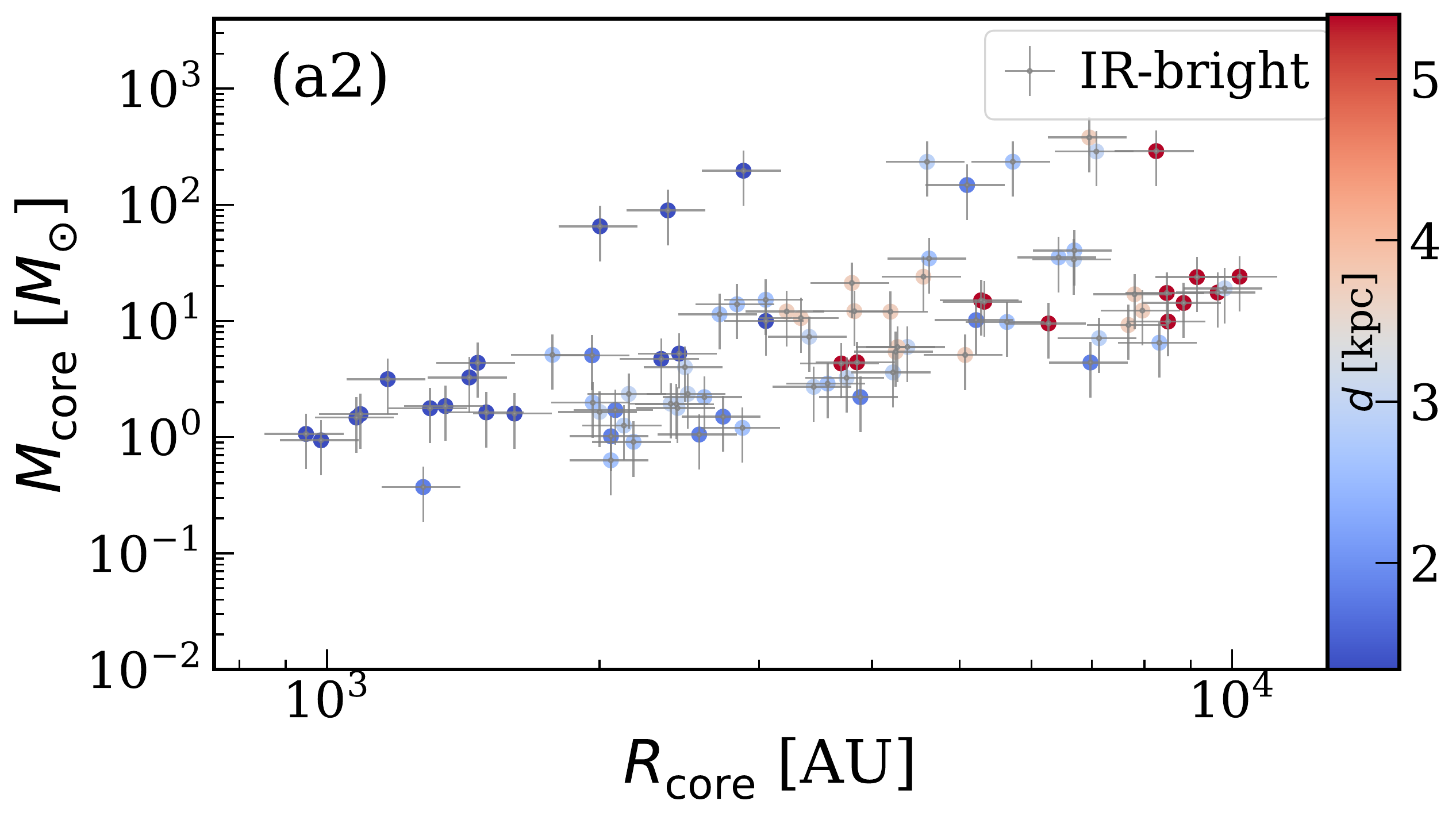}
\includegraphics[width=2.1 in]{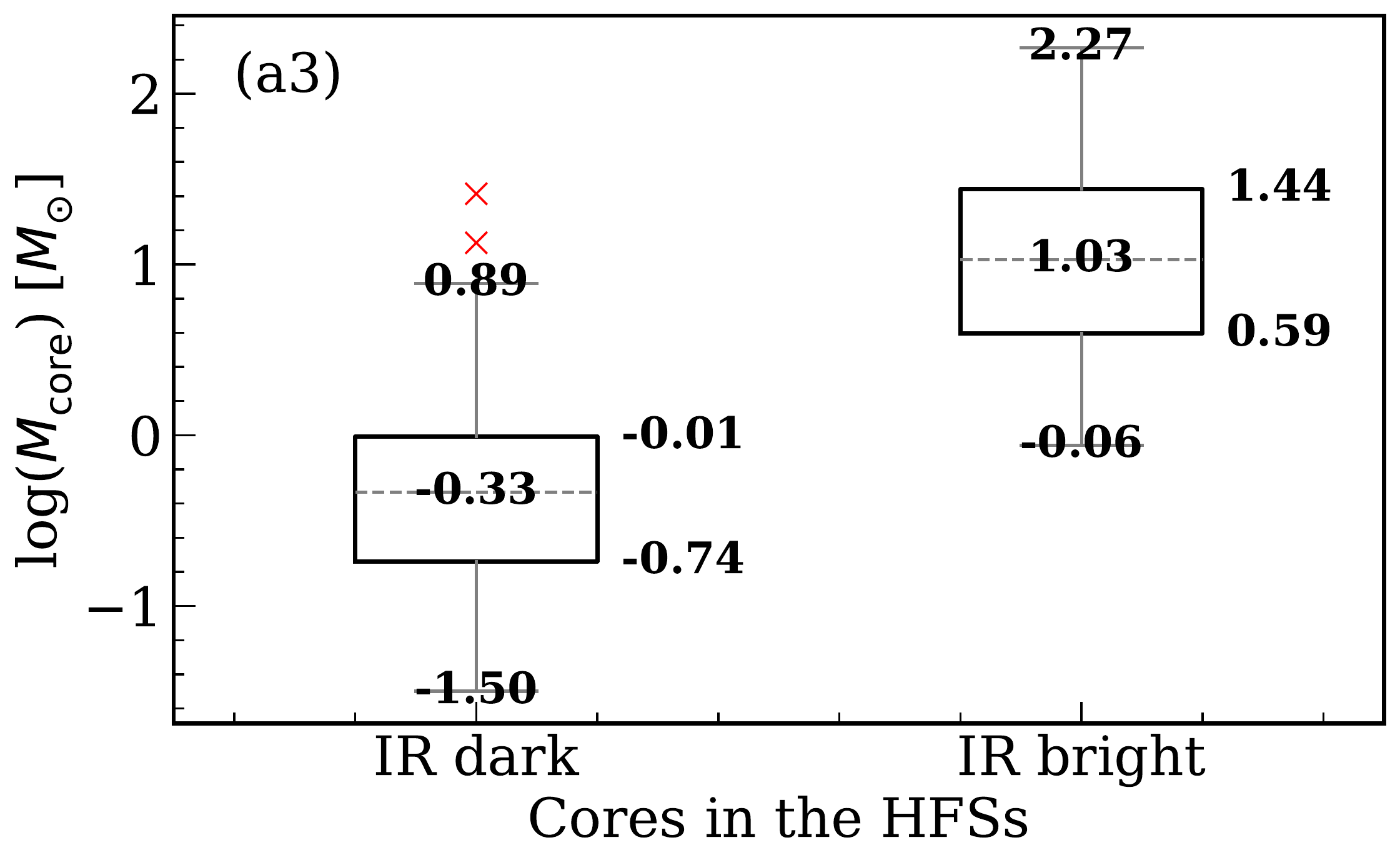}
\includegraphics[width=2.35 in]{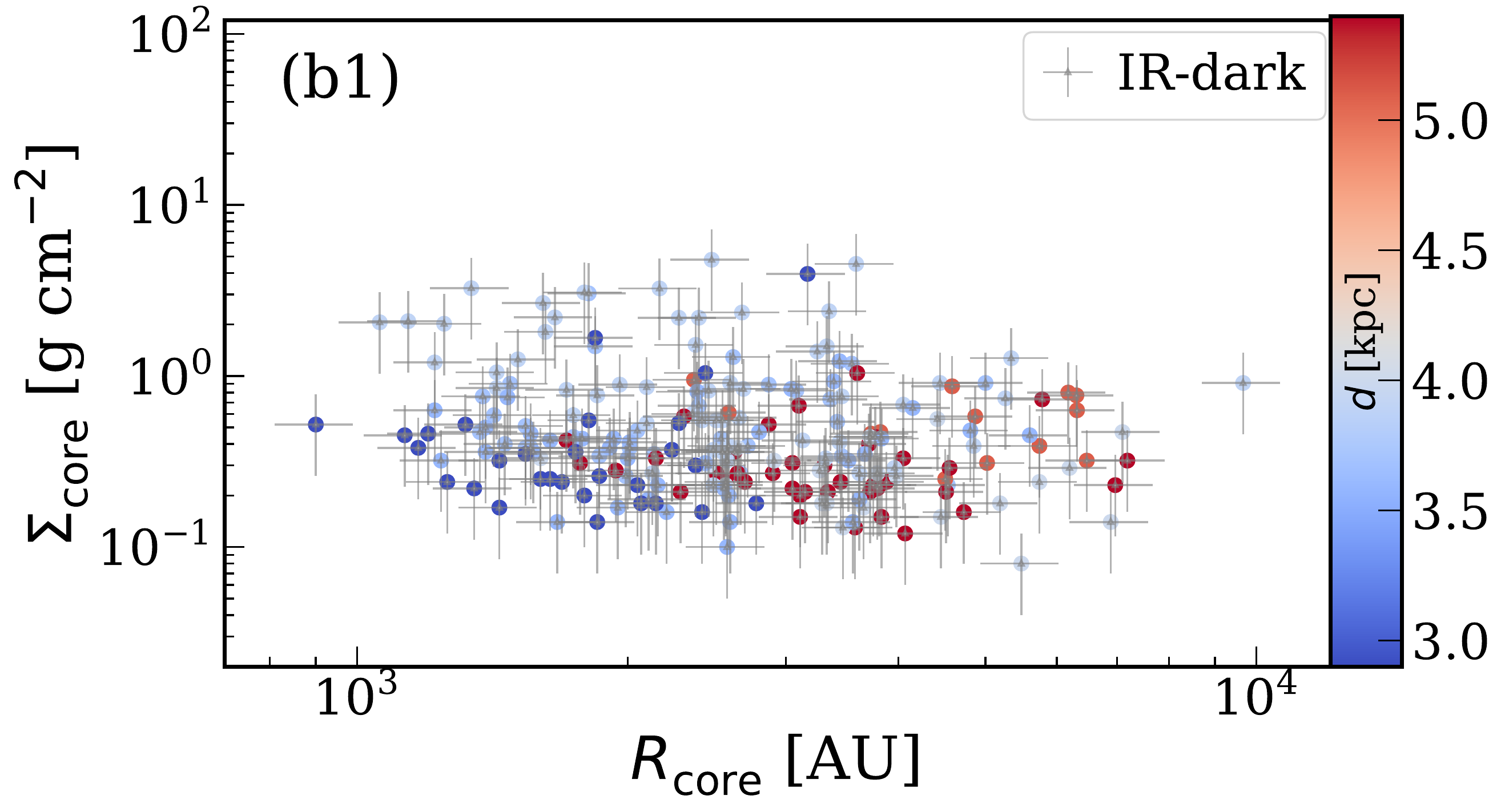}
\includegraphics[width=2.35 in]{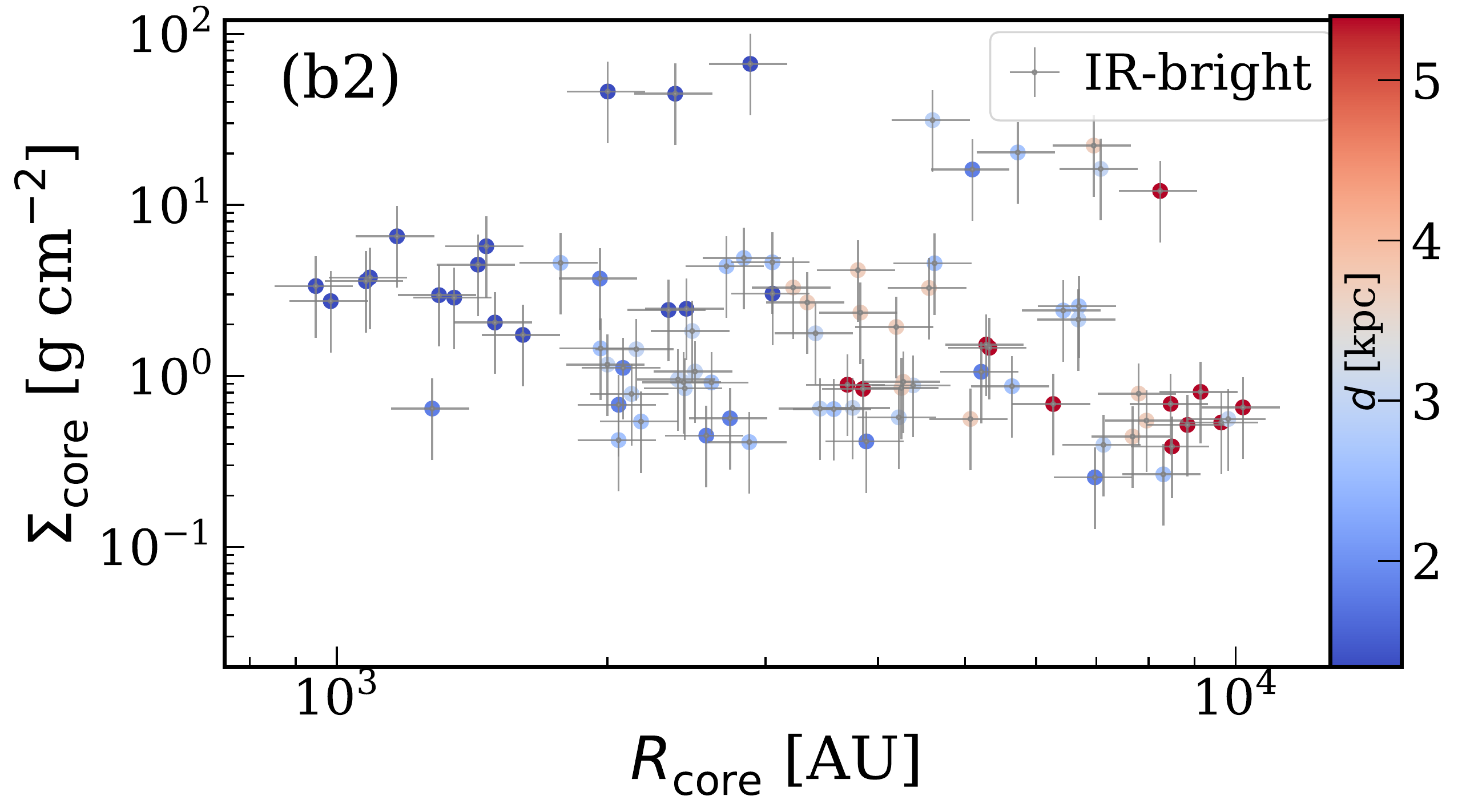}
\includegraphics[width=2.1 in]{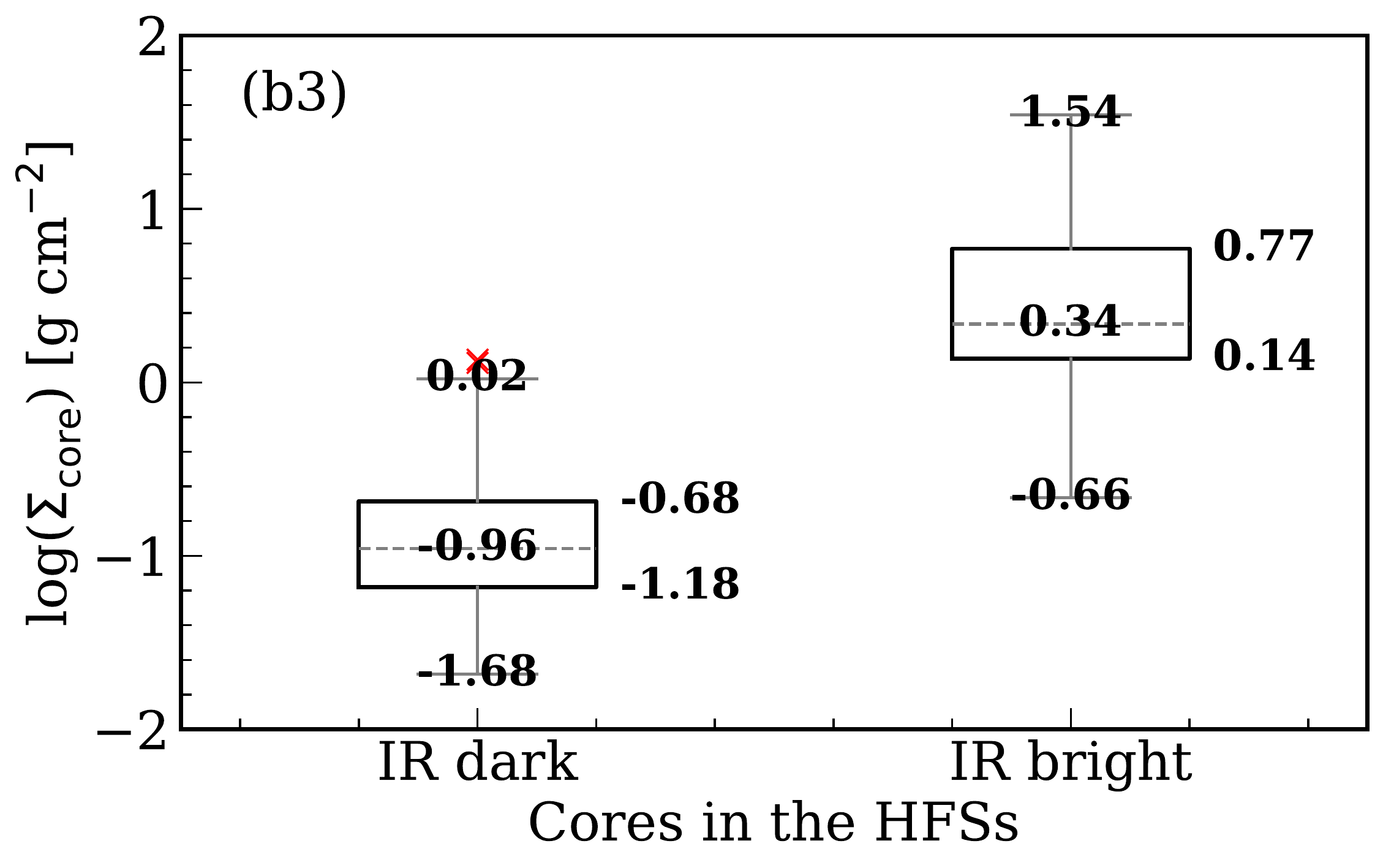}

\caption{Panels\,(a1--a3): mass ($M_{\rm core}$) distribution of cores.  Panels\,a1, and a2 show the $M_{\rm core}$ distribution against the radius of cores located in the IR-dark, and IR-bright HFSs, respectively.  The colors  of dots in both panels represent the distances of the cores.
Panel\,a3 displays a box-whisker plot summarising the $M_{\rm core}$ distribution of the cores in the two IR types of HFSs. Panels\,(b1--b3): same as Panels\,(a1--a3) but for the 
mass surface density ($\Sigma_{\rm core}$) distribution.  
In the box-whisker plots, the numbers associated with the boxes from the top to bottom represent the upper quartile, median (inside the box), and lower quartile, respectively. The red crosses indicate the outliers outside 1.5 times the interquartile range either 
above the upper quartile or below the lower quartile. 
}
\label{fig:core_mass_msig}
\end{figure*}

\begin{figure}
\centering
\includegraphics[width=3.3 in]{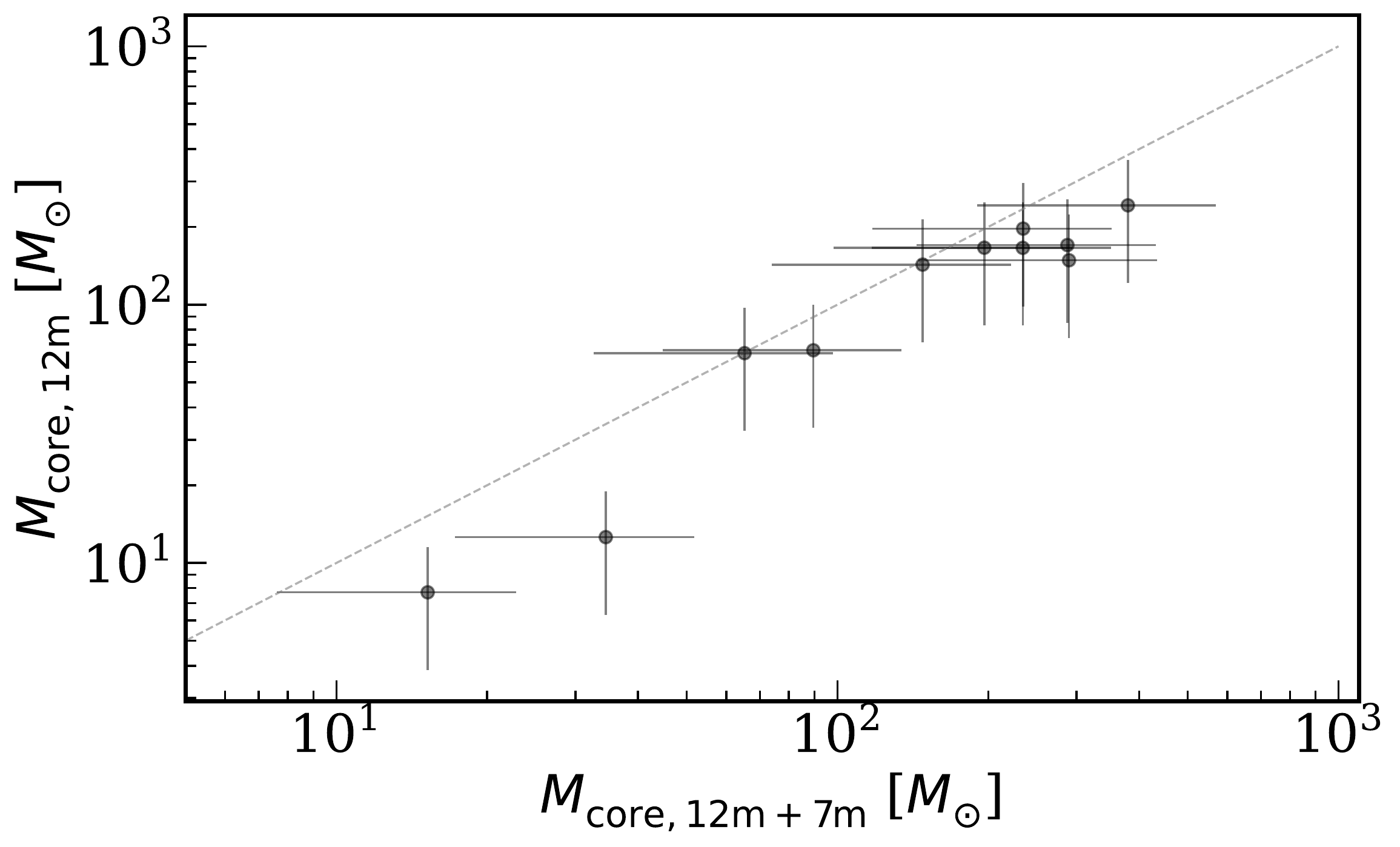}
\caption{Comparison between  determinations of the masses of the massive 
cores from the nine IR-bright HFSs, each having one clump.  The core masses were derived  first from the ATOMS 12m data alone, and  second the combined 12m+7m data.
The dashed line indicates  equal mass from each set of observations.
}
\label{fig:core_mass_msig2}
\end{figure}

\subsection{Clumps in the HFSs}
\label{sec:clumps}
Clumps are one of the characteristic hierarchical structures of the HFS clouds.
From Fig.\,\ref{fig:overview_hfs}, one can see that each HFS has a dominant central clump.
Generally, such density structures can be  identified using several widely-used algorithms, such as {\it Dendrogram} and {\it CASA-imfit}. However, the use of these algorithms is limited by the intensity contrast of the clumps  with respect to their natal clouds especially for the IR-dark HFSs studied here where the contrast is low (see Fig.\,\ref{fig:overview_hfs}). In these cases, the extracted clumps tend to have a large aspect ratio of $>3$ (i.e., ratio of the major to minor axis) indicative 
of more than one  entity within them. To avoid inaccurate clump identification for these low-contrast cases and to maintain uniformity, we define a circular aperture to enclose the enhanced 870\,\um\ emission in the central area of both IR-dark and bright HFSs analysed here.  From a careful scrutiny of the 870\,\um\ maps of the entire sample,
an optimum aperture with radius 0.25\,pc is found suitable to encompass most of the 870\,\um\ emission (see Fig.\,\ref{fig:overview_hfs_app}). Hence, we consider this as the radius of the clump, $R_{\rm clump}$.
Given this definition, the integrated flux ($F^{\rm int}_{\rm 870\mu m}$) of the central clumps in all HFSs were extracted from the ATLASGAL 870\,\um\ image.
In addition, we retrieved the dust temperature ($T_{\rm dust}$) of the clumps from \citet{San19} for those in the IR-dark HFSs and from 
\citet{Liu20a} for those in the IR-bright HFSs. 
With the above parameters, the clump mass ($M_{\rm clump}$) was
computed following Eqs.\,B1--B2 of \citet{Liu21}. In the computation, we assumed the gas-to-dust mass ratio to be $R_{\rm gd}=100$, and
the dust opacity per gram of dust to be $k_{\rm 870 \mu m}=1.78$\,cm$^{2}$~g$^{-1}$,
which corresponds to the opacity of dust grains with thin ice mantles at gas densities of $10^{6}$\,\pcmcu\ \citep{Oss94}. 
The mass surface density ($\Sigma_{\rm clump}$) of the clumps was derived from $\Sigma_{\rm clump} = M_{\rm clump}/(\pi R_{\rm clump}^2)$. 
The derived parameters are listed in Table\,\ref{tab:hfs_cloud}. 
According to \citet{San17, San19}, the uncertainty of both $M_{\rm clump}$ and $\Sigma_{\rm clump}$ could be about $50\%$, which accounts for  
the combined uncertainties from $k_{\rm 870 \mu m}$ ($\sim 30\%$), $R_{gd}$ ($\sim 20\%$), $T_{\rm d}$ ($\sim 20\%$), and the kinematic distance ($\sim10\%$).

The central clumps have a median $M_{\rm clump}$ of $\sim 223$\,\msun\ in a range of [153, 685]\,\msun\ in the IR-dark HFSs. For the IR-bright HFSs, the median $M_{\rm clump}$ is $\sim$649\,\msun\ in a range of [153, 975]\,\msun. $\Sigma_{\rm clump}$ has a median value of 0.24\,g\,\pcmsq\ in a range of [0.16, 0.73]\,g\,\pcmsq\ 
in the IR-dark HFSs, and 0.69\,g\,\pcmsq\ in a range of [0.16, 1.04]\,g\,\pcmsq\ in the IR-bright HFSs.
Overall, the estimated $\Sigma_{\rm clump}$ 
values in all HFSs studied here satisfy the empirical high-mass star formation threshold of $\Sigma_{\rm crit}\geq0.05$\,g~\pcmsq, which was
derived from the mass–-size relationship established using the ATLASGAL massive clumps 
containing high-mass star-forming signatures (e.g., methanol masers, and 
\hii~regions, \citealt{Urq14}). This provides evidence that the central clumps in both IR-dark and IR-bright HFSs are dense enough  to 
form high-mass stars. Evidence of high-mass star formation in the IR-dark HFSs (i.e., HFSs\,1--8)  has been 
suggested in  \citet{San19}, while the same  inference in the IR-bright HFSs (i.e., HFSs\,9--17) is strengthened by  
their associated high luminosities of $L_{\rm bol}\ga10^4$\,\lsun\ (see Table\,\ref{tab:hfs_cloud}). 

To quantitatively describe the evolutionary stage of clumps that can represent the stage of their natal HFSs in terms of high-mass star formation, 
we  consider the bolometric luminosity to mass ratio $L_{\rm bol}$/$M_{\rm clump}$, where the bolometric luminosity of each clump, $L_{\rm bol}$, can be found in 
Table\,\ref{tab:hfs_cloud}.  This is approximately taken to be the luminosity of their centrally located YSOs \citep{Bro96,Con13,Liu20a,Liu20b,Liu21} This ratio is  independent  of distance, and has been widely used as an indicator of the evolutionary stage of clumps \egcite{Guz15,Liu21}.
The clumps in IR-dark and bright HFSs have a median $L_{\rm bol}$/$M_{\rm clump}$ value of 0.14\,\lsun/\msun\ in a range of [0.04, 0.46]\,\lsun/\msun,
and a median value of 105\,\lsun/\msun\ in a range of [3.58, 204.62]\,\lsun/\msun, respectively. 
The ratio $L_{\rm bol}$/$M_{\rm clump}$ has been used as a diagnostic tool to probe the evolutionary stages of observed clumps \egcite{Urq14, Gia17,Eli22}. Based on the results discussed in these papers, values of $L_{\rm bol}$/$M_{\rm clump}$ $\la$ 2 have been associated with a very early evolutionary phase of mass accretion and possibly the beginning of protostellar activity. Whereas, ratios between 2--40 are shown to represent a later evolutionary phase where the protostar grows in mass with continuing accretion reaching the zero age main sequence around $L_{\rm bol}$/$M_{\rm clump} \sim 10$. Beyond a ratio of $\ga$ 40, onset of radio emission with detection of hypercompact and UCHII regions.
 This strongly supports  our conjecture that 
the ensemble of 8 IR-dark HFSs are at an earlier stage of high-mass star formation than that of the 9 IR-bright HFSs (see Sect.\,\ref{sec:sample}). 
Following this evolution,  an overall increasing trend of both $M_{\rm clump}$ and $\Sigma_{\rm clump}$ can be found from the IR-dark (223\,\msun\ and 
0.24\,g\,\pcmsq) to IR-bright (649\,\msun\ and 0.69\,g\,\pcmsq) stage of HFSs.

\begin{table*}
\centering
\caption{Parameters of cores in HFSs.}
\label{tab:cores}
\resizebox{10cm}{!}{
\begin{tabular}{ccllllllll}
\hline\hline
           HFS cloud &              Core ID &                   RA &                  DEC &  $R_{\rm core}$ &  $R_{\rm core}$ & $F_{\rm cont}^{int.}$ &  $M_{\rm core}$ & $\Sigma_{\rm core}$ &          Assoc.   \\
                 &                      &                J2000  &  J2000 & $\arcsec$ &     AU &    mJy & $M_{\odot}$ & g cm$^{-2}$ &                        \\

\hline
  HFS9  &     ALMA1  &           13:51:58.35  &           -61:15:41.9  &    1.53  &    8242  &   56.63  &   289.9  &   12.07  &                   1,2  \\ 
  HFS9  &     ALMA2  &           13:51:55.31  &           -61:16:05.4  &    1.57  &    8467  &    3.40  &    17.4  &    0.69  &                     0  \\ 
  HFS9  &     ALMA3  &           13:52:00.74  &           -61:15:53.2  &    1.89  &   10190  &    4.70  &    24.1  &    0.66  &                     0  \\ 
  HFS9  &     ALMA4  &           13:51:57.10  &           -61:15:51.0  &    1.69  &    9143  &    4.66  &    23.9  &    0.81  &                     4  \\ 
  HFS9  &     ALMA5  &           13:51:58.09  &           -61:15:38.8  &    1.79  &    9639  &    3.43  &    17.6  &    0.53  &                     0  \\ 
  HFS9  &     ALMA6  &           13:51:58.58  &           -61:15:33.9  &    0.98  &    5280  &    2.94  &    15.0  &    1.53  &                     4  \\ 
  HFS9  &     ALMA7  &           13:51:57.87  &           -61:15:31.8  &    0.99  &    5323  &    2.86  &    14.6  &    1.46  &                     4  \\ 
  HFS9  &     ALMA8  &           13:52:01.76  &           -61:15:25.0  &    1.16  &    6267  &    1.86  &     9.5  &    0.69  &                     0  \\ 
  HFS9  &     ALMA9  &           13:51:57.56  &           -61:16:06.4  &    0.71  &    3851  &    0.86  &     4.4  &    0.84  &                     4  \\ 
  HFS9  &    ALMA10  &           13:52:03.09  &           -61:15:53.5  &    0.69  &    3700  &    0.84  &     4.3  &    0.89  &                     0  \\ 
  HFS9  &    ALMA11  &           13:51:58.29  &           -61:15:45.9  &    1.57  &    8496  &    1.93  &     9.9  &    0.39  &                     4  \\ 
  HFS9  &    ALMA12  &           13:51:59.32  &           -61:15:41.1  &    1.64  &    8839  &    2.79  &    14.3  &    0.52  &                     4  \\ 
 HFS10  &     ALMA1  &           15:43:18.94  &           -54:07:35.6  &    1.08  &    1962  &    8.69  &     5.1  &    3.71  &                   2,4  \\ 
 HFS10  &     ALMA2  &           15:43:16.62  &           -54:07:14.8  &    2.80  &    5095  &  254.36  &   147.9  &   16.12  &                   2,4  \\ 
 HFS10  &     ALMA3  &           15:43:18.36  &           -54:07:34.1  &    2.87  &    5213  &   17.49  &    10.2  &    1.06  &                     0  \\ 
 HFS10  &     ALMA4  &           15:43:14.23  &           -54:07:27.7  &    1.13  &    2059  &    1.75  &     1.0  &    0.68  &                     0  \\ 
 HFS10  &     ALMA5  &           15:43:16.91  &           -54:07:22.4  &    3.83  &    6973  &    7.55  &     4.4  &    0.26  &                     0  \\ 
 HFS10  &     ALMA6  &           15:43:16.07  &           -54:07:10.9  &    2.13  &    3884  &    3.80  &     2.2  &    0.41  &                     0  \\ 
 HFS10  &     ALMA7  &           15:43:17.15  &           -54:07:07.5  &    1.51  &    2739  &    2.58  &     1.5  &    0.57  &                     0  \\ 
 HFS10  &     ALMA8  &           15:43:16.96  &           -54:06:59.4  &    1.14  &    2082  &    2.94  &     1.7  &    1.12  &                     0  \\ 
 HFS10  &     ALMA9  &           15:43:15.76  &           -54:07:22.7  &    1.42  &    2577  &    1.81  &     1.1  &    0.45  &                     0  \\ 
 HFS10  &    ALMA10  &           15:43:18.49  &           -54:07:17.7  &    0.70  &    1277  &    0.64  &     0.4  &    0.64  &                     0  \\ 
 HFS11  &     ALMA1  &           15:55:48.46  &           -52:43:07.9  &    2.16  &    5723  &  190.43  &   234.7  &   20.27  &                   1,3  \\ 
 HFS11  &     ALMA2  &           15:55:46.39  &           -52:43:23.5  &    1.07  &    2838  &   11.30  &    13.9  &    4.89  &                     4  \\ 
 HFS11  &     ALMA3  &           15:55:47.93  &           -52:43:13.8  &    2.53  &    6693  &   32.84  &    40.5  &    2.56  &                     0  \\ 
 HFS11  &     ALMA4  &           15:55:48.99  &           -52:43:12.0  &    1.02  &    2714  &    9.26  &    11.4  &    4.38  &                     0  \\ 
 HFS11  &     ALMA5  &           15:55:48.87  &           -52:43:02.9  &    2.43  &    6430  &   28.63  &    35.3  &    2.41  &                     0  \\ 
 HFS11  &     ALMA6  &           15:55:46.18  &           -52:43:00.7  &    0.67  &    1774  &    4.14  &     5.1  &    4.58  &                     0  \\ 
 HFS12  &     ALMA1  &           16:30:58.82  &           -48:43:53.9  &    1.58  &    4601  &  156.58  &   234.3  &   31.30  &                   1,2  \\ 
 HFS12  &     ALMA2  &           16:30:58.61  &           -48:43:57.4  &    2.44  &    7128  &    4.75  &     7.1  &    0.40  &                     0  \\ 
 HFS12  &     ALMA3  &           16:30:59.26  &           -48:43:53.0  &    1.45  &    4219  &    2.41  &     3.6  &    0.57  &                     0  \\ 
 HFS12  &     ALMA4  &           16:30:57.17  &           -48:43:54.4  &    1.18  &    3449  &    1.81  &     2.7  &    0.64  &                     0  \\ 
 HFS12  &     ALMA5  &           16:30:57.37  &           -48:43:39.6  &    2.29  &    6685  &   22.57  &    33.8  &    2.14  &                   2,4  \\ 
 HFS12  &     ALMA6  &           16:31:01.53  &           -48:44:05.2  &    0.73  &    2128  &    0.84  &     1.3  &    0.79  &                     0  \\ 
 HFS12  &     ALMA7  &           16:30:59.32  &           -48:43:51.1  &    0.84  &    2439  &    1.19  &     1.8  &    0.85  &                     0  \\ 
 HFS13  &     ALMA1  &           16:38:50.55  &           -47:28:01.8  &    2.34  &    7078  &  179.71  &   287.7  &   16.24  &                   1,3  \\ 
 HFS13  &     ALMA2  &           16:38:51.68  &           -47:28:20.2  &    0.71  &    2155  &    1.47  &     2.4  &    1.43  &                     4  \\ 
 HFS13  &     ALMA3  &           16:38:51.24  &           -47:28:13.9  &    1.13  &    3410  &    4.56  &     7.3  &    1.78  &                     4  \\ 
 HFS13  &     ALMA4  &           16:38:50.06  &           -47:28:04.2  &    3.25  &    9812  &   11.91  &    19.1  &    0.56  &                     4  \\ 
 HFS13  &     ALMA5  &           16:38:49.70  &           -47:27:55.7  &    0.82  &    2485  &    2.50  &     4.0  &    1.83  &                     4  \\ 
 HFS13  &     ALMA6  &           16:38:50.61  &           -47:27:52.4  &    1.45  &    4377  &    3.73  &     6.0  &    0.88  &                     0  \\ 
 HFS13  &     ALMA7  &           16:38:51.42  &           -47:27:48.1  &    1.24  &    3750  &    2.02  &     3.2  &    0.65  &                     4  \\ 
 HFS13  &     ALMA8  &           16:38:51.77  &           -47:28:10.0  &    0.81  &    2432  &    1.20  &     1.9  &    0.92  &                     4  \\ 
 HFS13  &     ALMA9  &           16:38:53.98  &           -47:28:02.8  &    0.66  &    2000  &    1.03  &     1.6  &    1.16  &                     0  \\ 
 HFS13  &    ALMA10  &           16:38:51.60  &           -47:27:56.2  &    0.79  &    2397  &    1.21  &     1.9  &    0.95  &                     4  \\ 
 HFS13  &    ALMA11  &           16:38:50.28  &           -47:27:47.3  &    0.83  &    2504  &    1.47  &     2.4  &    1.06  &                     0  \\ 
 HFS14  &     ALMA1  &           16:46:06.08  &           -45:36:43.2  &    1.76  &    4625  &   28.41  &    34.5  &    4.56  &                   1,2  \\ 
 HFS14  &     ALMA2  &           16:46:07.30  &           -45:36:40.9  &    1.16  &    3053  &   12.53  &    15.2  &    4.62  &                     4  \\ 
 HFS14  &     ALMA3  &           16:46:06.88  &           -45:36:52.2  &    2.15  &    5639  &    8.07  &     9.8  &    0.87  &                     4  \\ 
 HFS14  &     ALMA4  &           16:46:04.59  &           -45:36:53.9  &    1.09  &    2877  &    0.99  &     1.2  &    0.41  &                     0  \\ 
 HFS14  &     ALMA5  &           16:46:05.99  &           -45:36:51.5  &    0.75  &    1966  &    1.63  &     2.0  &    1.45  &                     4  \\ 
 HFS14  &     ALMA6  &           16:46:05.72  &           -45:36:48.3  &    0.99  &    2611  &    1.82  &     2.2  &    0.92  &                     4  \\ 
 HFS14  &     ALMA7  &           16:46:05.04  &           -45:36:47.1  &    0.78  &    2059  &    0.52  &     0.6  &    0.42  &                     4  \\ 
 HFS14  &     ALMA8  &           16:46:07.65  &           -45:36:39.1  &    1.36  &    3573  &    2.38  &     2.9  &    0.64  &                     0  \\ 
 HFS14  &     ALMA9  &           16:46:08.43  &           -45:36:36.5  &    0.83  &    2181  &    0.75  &     0.9  &    0.54  &                     0  \\ 
 HFS14  &    ALMA10  &           16:46:06.55  &           -45:36:39.9  &    3.16  &    8309  &    5.35  &     6.5  &    0.27  &                     4  \\ 
 HFS15  &     ALMA1  &           17:05:11.15  &           -41:29:07.0  &    1.74  &    2380  &  271.70  &    89.5  &   44.69  &                   3,4  \\ 
 HFS15  &     ALMA2  &           17:05:10.23  &           -41:29:33.0  &    1.07  &    1467  &   13.24  &     4.4  &    5.73  &                     0  \\ 
 HFS15  &     ALMA3  &           17:05:09.52  &           -41:29:06.3  &    0.85  &    1167  &    9.58  &     3.2  &    6.55  &                     0  \\ 
 HFS15  &     ALMA4  &           17:05:11.84  &           -41:29:01.3  &    0.95  &    1300  &    5.38  &     1.8  &    2.97  &                     0  \\ 
 HFS15  &     ALMA5  &           17:05:09.57  &           -41:28:57.4  &    0.99  &    1351  &    5.62  &     1.9  &    2.87  &                     0  \\ 
 HFS15  &     ALMA6  &           17:05:10.92  &           -41:28:47.3  &    1.71  &    2340  &   14.29  &     4.7  &    2.43  &                     0  \\ 
 HFS15  &     ALMA7  &           17:05:09.40  &           -41:29:28.1  &    0.79  &    1078  &    4.47  &     1.5  &    3.59  &                     0  \\ 
 HFS15  &     ALMA8  &           17:05:12.17  &           -41:29:08.5  &    0.80  &    1089  &    4.79  &     1.6  &    3.76  &                     0  \\ 
 HFS16  &     ALMA1  &           17:26:42.56  &           -36:09:18.2  &    2.15  &    2885  &  623.10  &   196.4  &   66.75  &                   1,3  \\ 
 HFS16  &     ALMA2  &           17:26:43.61  &           -36:09:17.1  &    1.50  &    2002  &  207.04  &    65.3  &   46.03  &                     4  \\ 
 HFS16  &     ALMA3  &           17:26:42.96  &           -36:09:40.0  &    0.74  &     985  &    2.98  &     0.9  &    2.74  &                     4  \\ 
 HFS16  &     ALMA4  &           17:26:40.26  &           -36:09:36.0  &    1.20  &    1611  &    5.06  &     1.6  &    1.74  &                     0  \\ 
 HFS16  &     ALMA5  &           17:26:45.12  &           -36:09:19.8  &    0.71  &     948  &    3.38  &     1.1  &    3.35  &                     0  \\ 
 HFS16  &     ALMA6  &           17:26:42.95  &           -36:09:13.3  &    2.28  &    3054  &   31.69  &    10.0  &    3.03  &                   2,4  \\ 
 HFS16  &     ALMA7  &           17:26:43.59  &           -36:09:09.3  &    1.07  &    1436  &   10.35  &     3.3  &    4.47  &                     4  \\ 
 HFS16  &     ALMA8  &           17:26:42.12  &           -36:09:20.7  &    1.12  &    1500  &    5.18  &     1.6  &    2.05  &                     0  \\ 
 HFS16  &     ALMA9  &           17:26:43.38  &           -36:09:19.6  &    1.83  &    2449  &   16.63  &     5.2  &    2.47  &                     0  \\ 
 HFS17  &     ALMA1  &           18:53:18.64  &            +1:24:46.5  &    1.03  &    3801  &    8.84  &    21.2  &    4.16  &                   3,4  \\ 
 HFS17  &     ALMA2  &           18:53:18.07  &            +1:25:25.3  &    1.88  &    6952  &  158.09  &   379.9  &   22.23  &                   1,2  \\ 
 HFS17  &     ALMA3  &           18:53:18.41  &            +1:24:54.2  &    1.23  &    4559  &   10.00  &    24.0  &    3.27  &                     4  \\ 
 HFS17  &     ALMA4  &           18:53:18.33  &            +1:25:13.2  &    0.90  &    3338  &    4.41  &    10.6  &    2.69  &                     4  \\ 
 HFS17  &     ALMA5  &           18:53:18.53  &            +1:25:18.1  &    0.87  &    3220  &    5.02  &    12.1  &    3.29  &                     4  \\ 
 HFS17  &     ALMA6  &           18:53:18.23  &            +1:25:19.9  &    2.08  &    7680  &    3.84  &     9.2  &    0.44  &                   2,4  \\ 
 HFS17  &     ALMA7  &           18:53:18.07  &            +1:25:30.1  &    2.15  &    7958  &    5.11  &    12.3  &    0.55  &                     0  \\ 
 HFS17  &     ALMA8  &           18:53:18.66  &            +1:25:27.5  &    1.13  &    4193  &    5.00  &    12.0  &    1.93  &                     4  \\ 
 HFS17  &     ALMA9  &           18:53:18.51  &            +1:25:37.3  &    2.11  &    7801  &    7.06  &    17.0  &    0.79  &                     0  \\ 
 HFS17  &    ALMA10  &           18:53:18.75  &            +1:26:00.4  &    1.03  &    3824  &    5.05  &    12.1  &    2.35  &                     4  \\ 
 HFS17  &    ALMA11  &           18:53:18.72  &            +1:24:44.1  &    1.15  &    4248  &    2.26  &     5.4  &    0.85  &                     0  \\ 
 HFS17  &    ALMA12  &           18:53:18.43  &            +1:24:46.2  &    1.37  &    5070  &    2.12  &     5.1  &    0.56  &                     0  \\ 
 HFS17  &    ALMA13  &           18:53:18.34  &            +1:25:09.8  &    1.15  &    4269  &    2.48  &     6.0  &    0.92  &                     0  \\ 

\hline
\end{tabular}
}

\begin{flushleft}
{\bf Note:} This table  includes only the parameters of cores in the nine IR-bright HFSs, while the same parameters of the cores
in the eight IR-dark HFSs are referred to tables\,3, and 4 of \citet{San19}. 
$F_{\rm cont}^{\rm int.}$ represents the integrated 3\,mm continuum flux of the cores. Core association ranges from 0 to 4, where
0 = prestellar candidate, 1 = molecular outflow, 2 = hot core, 3 = compact \hii\ region, and 4 = point-like 8/24\,\um\ object.
\end{flushleft}
\end{table*}

\subsection{Cores in the HFSs}
\label{sec:cores}

\subsubsection{Core identification}
\label{sec:core_identify}
The high-resolution ($1.2$\arcsec--2\arcsec) ALMA continuum data, that correspond to linear scales 0.02 and 0.03\,pc at the typical distances of the IR-dark and bright HFSs, respectively, enable the identification of compact cores where stars could form.
 A total of 224 compact cores in the 8 IR-dark HFSs have already been identified from ASHES 1.3\,mm continuum by \citet{San19}. Slightly different approaches have been implemented in \citet{San19} and \citet{Liu21} to identify compact cores. While the former study used the {\it Dendrogram} algorithm, the later used a two-step process in which the initial identification was carried out using {\it Dendrogram} then followed by CASA--{\it imfit} to estimate the parameters. Examining the performance of both schemes (especially for the low-mass cores in the IR-dark HFSs) and to maintain uniformity, we use {\it Dendrogram} alone to extract cores in the 9 IR-bright HFSs from ATOMS 3\,mm, combined 12m+7m continuum data. It is worth mentioning here that there are a suite of clump/core identification algorithms available (e.g., {\it Clumpfind} by \citealt{Wil94}; {\it getsf} by \citealt{Men21}) and {\it Dendrogram} is one such robust algorithm widely used in similar studies \egcite{Ros08,Gin16,Off22}. While a comparative study, which is beyond the focus of this paper, would help highlight the nuances of each, the overall interpretation based on the retrieved parameters would remain the same.

Following \citet{San19}, an intensity threshold of 2.5\,$\sigma$,  a step of 1.0\,$\sigma$ (the rms noise of the continuum data, see Sect.\,\ref{sec:almadata}), 
and a minimum number of pixels equal to those contained in each synthesized beam (half of the beam considered in \citealt{San19}) were used as input parameters. The algorithm identifies the small structures, called "leaves", which cannot further break up into smaller structures, and are thus defined here as cores. 
Finally, cores with integrated flux densities less than 4\,$\sigma$ were  excluded to avoid spurious identification, where 4\,$\sigma$ was determined
from the corresponding negative level (i.e., $-4$\,$\sigma$) within which the interferometric sidelobe effects cannot be ruled out.
A total of 86 compact cores are obtained in the 9 IR-bright HFSs.
The parameters retrieved from the {\it Dendrogram} analysis are listed in
Table\,\ref{tab:cores}. These include the
core coordinates, radius ($R_{\rm core}$),  
 and integrated flux ($F_{\rm cont}^{\rm int}$).
The same parameters for the compact cores in the 8 IR-dark HFSs are referred to Tables\,3 and 4 of \citet{San19}.

 The number of the cores in the IR-bright HFSs found  in this study is about four times greater than that reported in \citet{Liu21} for the same regions.
Apart from the slightly different parameter used, a possible reason  is the use of the higher resolution 12m data for core extraction in \citet{Liu21} as opposed to the combined 12m+7m data used here, which will be discussed below. 
Similar differences were noted by \citet{San19} where inclusion of more extended emission with the 7m array leads to detection  of 20\% greater number of cores.

\subsubsection{Star-forming nature of cores}
\label{sec:core_nature}
The star-forming  activity in cores can be inferred from YSO signatures such as outflows, hot cores, compact \hii\ regions. 
\citet{Li 20} provided the catalogues of CO and SiO outflows associated with the cores in the IR-dark HFSs while the  association of 
hot cores and \hii\ regions  with the cores in the IR-bright HFSs  are tabulated in 
\citet{Liu21}, who complied these based on the detection of rich complex organic molecular lines, and \halpha\ line.
Presence of point-like 24\,\um\ or slightly extended but compact 8\,\um\ sources, which are likely to be candidate YSOs and hence could be considered as signposts of star formation.
These associations are listed in the last column of Table\,\ref{tab:cores}.
In total, 112 out of 310 cores are protostellar ones associated with one or more star-forming signatures (see above). The remaining 198 cores without any of these signatures are treated here as prestellar candidates. 

\subsubsection{Derived parameters of cores}
\label{sec:core_params_derived}
From the parameters (e.g., radius $R_{\rm core}$, and integrated flux $F_{\rm cont.}^{\rm int.}$) of the cores derived from the {\it Dendrogram} analysis discussed in Sect.\,\ref{sec:core_identify}, 
other parameters such as mass and mass surface densities can be estimated.
Cores without detectable star-forming signatures are roughly assigned the temperature of their natal clumps.
For cores with associated star-forming signatures, the temperature was assumed to be 50\,K for those associated with outflows \citep{San19}, and 100\,K for those
associated either with hot cores or compact \hii\ regions \citep{Liu21}. 
 The mass ($M_{\rm core}$) and mass surface density ($\Sigma_{\rm core}$) parameters of the cores were calculated following the same approach outlined 
in Sect.\,\ref{sec:clumps}, where the dust opacities per gram of dust were adjusted to be 0.9\,cm$^{2}$~g$^{-1}$ for 1\,mm and  0.2\,cm$^{2}$~g$^{-1}$ for 3\,mm 
according to \citet{Oss94}. The uncertainties of $M_{\rm core}$ and $\Sigma_{\rm core}$ are around $\sim 50\%$ (see Sect.\,\ref{sec:clumps}).
The derived physical parameters are listed in Table\,\ref{tab:cores}.

It is worth noting that for those five cores associated with compact \hii\ regions (see Table\,\ref{tab:cores}) 
that lie in five of nine IR-bright HFSs (i.e., HFSs\,11, 13, 15, 16, 17), 
the continuum flux of cores would have contribution from both dust thermal emission and free-free emission. The values tabulated have been subtracted for the free-free emission component.
To estimate the contribution of free-free emission, we used the H40$\alpha$ hydrogen recombination line observations under the assumption of local thermodynamical equilibrium and optically thin emission. The free-free emission intensity was estimated
via the following relation \egcite{Mot22}:
\begin{equation}\label{eq:freefree}
S_{\rm ff} =  1.43\times 10^{-4} S_\mathrm{RRL} [\frac{\nu}{\rm GHz}]^{-1.1} [\frac{T_e}{\rm K}]^{1.15} (1 + \frac{N_\mathrm{He}}{N_\mathrm{H}})^{-1},
\end{equation}
where $S_\mathrm{RRL}$ is the integrated intensity of H$40\alpha$ over its velocity extent;
 $\nu = 99.0$\,GHz is the rest frequency of H40$\alpha$ observed in the ATOMS data; 
and we assume the electron temperature of $T_e = 6000$~K as well as a relative abundance of helium to hydrogen of $N_\mathrm{He}/N_\mathrm{H} = 0.08$. 
We assume an upper limit of $T_e = 6000$\,K to avoid over subtraction of free-free continuum for the small scale extracted cores \citep{Liu15}. 
The above estimated free-free emission intensity was subsequently subtracted from the 3\,mm continuum image to yield the dust continuum emission image.

Figure\,\ref{fig:core_mass_msig}\,(a--b) show the distribution of $M_{\rm core}$ and  $\Sigma_{\rm core}$ as a function of $R_{\rm core}$ for the cores
in both IR-dark and bright HFSs. The colors in dots in the figure indicate the distance distribution. 
The cores in both IR types of HFSs have a similar radius range of [0.9, 10]$\times10^3$\,AU, with an average radius ratio of $\sim1.4$ of 
cores in the IR-dark HFSs to those in  the IR-bright HFSs. 
As suggested by \citealt{Lou21}, this can be understood since the similar typical spatial resolutions of the ALMA observations toward the two IR-type HFSs  can yield the close derived sizes of the extracted cores.

Considering the median values of the mass and mass surface density in the IR-bright HFSs (5.7\,\msun\ and 1.4\,g\,\pcmsq) and the IR-dark HFSs (1.0\,\msun\ and 0.4\,g\,\pcmsq), it is seen that the mass and mass surface density is higher by factors of 6 and 3, respectively in the IR-bright HFSs (Fig.\,\ref{fig:core_mass_msig}\,c).
 From Fig.\,\ref{fig:core_mass_msig}\,(a--b), this difference can be seen to be independent of the source distance.
Instead, it could be in part a result of the observation bias caused by the different maximum recoverable scales (MRS) in the ASHES ($\sim 20$\arcsec) and ATOMS ($\sim60$\arcsec) combined 12m+7m data. As suggested in \citet{San19}, the higher MRS of the ATOMS combined data could intrinsically lead to higher fluxes and thus flux-derived parameters (e.g., mass and mass surface densities) of cores.  Different mass sensitivity of the surveys could also contribute since in the higher sensitivity ASHES data, more low-mass cores are detected. 
However, it is worth noting that this observation bias could not affect significantly  
the most massive cores (i.e., nine cores with the highest $M_{\rm core}$ and  $\Sigma_{\rm core}$ values in panels\,a--b), as can be
seen from Fig.\,\ref{fig:core_mass_msig2}.  Hence, in the analysis that follows, we focus on the most massive cores
in the IR-bright sample of HFSs. In Fig.\,\ref{fig:core_mass_msig2}, we compare the masses estimated from
the ATOMS 12m continuum data alone with a MRS of $\sim 18$\,\arcsec  (similar to the MRS of the ASHES data) with that derived from the 12m+7m combined data
having a MRS of $\sim 60$\arcsec. From this comparison, we estimate a nominal factor of $\sim $1.3 on average higher mass estimates for the cores extracted in the IR-bright HFSs using ATOMS 12m+7m data.
 As is also seen in the figure, the most massive cores in the IR-bright HFSs can be clearly distinguished from the majority of low mass cores in the distributions 
of both $M_{\rm core}$ and $\Sigma_{\rm core}$ (see Fig.\,\ref{fig:core_mass_msig}\,a2 and b2).
This trend could be either a consequence of the limited HFS sample investigated here or the evolutionary phase of the IR-bright HFSs, the latter being related to their preferred central location and sufficiently long accretion history.
Certainly, larger sample of such IR-dark and IR-bright HFSs are required
to examine the above possibilities.

\subsection{YSOs in the HFSs}
Investigating the spatial distribution of associated YSOs helps in understanding star formation in HFSs.
From the 24\,\um\ point source catalogue of the Galactic plane from Spitzer/MIPSGAL \citep{Gut15}, 
we searched for point sources associated with the HFS sample studied here, and obtained their 24\,\um\ photometric fluxes. 
Here, we have used different search areas depending on the spatial extent of the HFSs in the 8\,\um\ image (i.e., image size in Fig.\,\ref{fig:overview_hfs},  1.5\arcmin--4.8\arcmin).
In addition to the catalogued sources, we identified ten bright, point-like sources in the 24\,\um\ MIPSGAL images that were not included in the MIPSGAL point source catalogue. These were identified in the HFSs\,9, 11, 13, 14, 15, 16, and 17. For these sources, the photometry was performed manually using appropriate circular annulus whose inner and outer radii can represent well the point-like source and its associated background.

Since the typical distance of the HFSs studied here is $\sim3.6$\,kpc, some of the 24\,\um\ point sources obtained above could be foreground/background stars along the line of sight of the HFSs. 
To alleviate this, we correlate the identified sources with Gaia EDR3 data within a radius of $2$\arcsec. 
The parallax distances of the Gaia-matched sources were subsequently calculated. 
As a rough approximation, we assume that
the sources with distance estimates more than 10\% of the nearest HFS's distance are foreground/background stars. It is to be noted
that only a few sources were filtered out as foreground/background stars. Finally, 175 point sources that are likely associated with the HFSs studied here are retained for further analysis. 

Furthermore, we classify the identified YSOs cross-matching (within $2$\arcsec\ radius) using the SPICY catalogue that compiles $\sim$~120,000 Spitzer/IRAC candidate YSOs  for the Inner Galactic Midplane \citep{Kuh21}. 
This catalogue contains five classes of YSOs, including Class\,I, Flat-spectrum, Class\,II, Class\,III, 
and ``Uncertain'' YSOs that cannot be placed in any of the above four  classes. 
Class\,I YSOs are highly embedded protostars with the bolometric luminosity dominated by a spherical infalling envelope;
Class\,II YSOs are young stars surrounded by a substantial accreting disk; Class\,III YSOs
are young stars with most of their disk mass being dissipated. Flat spectrum YSOs are those in between Class\,I and Class\,II.
From the above exercise, 122 of the 175  detected 24\,\um\ sources are classified as YSOs (29 and 93 in the IR-dark and IR-bright HFSs, respectively) and 53 as ``unknown'' sources. 
To confirm the nature of the ``unknown'' sources 
requires detailed investigation of the photometric information over multiple wavelengths, which is beyond the scope of this work.

The YSO luminosity can be directly linked to the star-formation process,
i.e., either low or high-mass star formation.
In previous studies, the 24\,\um\ photometric flux of both low and high-mass protostellar sources was found to correlate well with their bolometric luminosity ($L_{\rm bol}$, e.g., \citealt{Dun08,Rag12}). 
Applying this empirical correlation, we estimated the luminosities of the 175 candidate YSOs from their 24\,\um\ fluxes.
Except for the one luminous YSO ($L_{\rm bol}\sim 10^4$\,\lsun; see Fig.\,\ref{fig:yso_distr}), the candidate YSOs in the IR-dark HFSs have low luminosities of $L_{\rm bol}<10^3$\,\lsun\ .  
Note that the location of the one luminous YSO is more than 1\,pc from the central hub of HFS\,3, and thus do not conflict with the classification of the host HFS as IR-dark type. 
In case of the IR-bright HFSs, 
the majority of the candidate YSOs are also found to have low luminosities though there are 12 high-luminosity sources found with $L_{\rm bol}\sim$~[$10^4$, $10^5$]\,\lsun.
The presence of a significant population of high-luminosity YSOs in the IR-bright HFSs agrees well with their more evolved stage inferred earlier.

\begin{figure*}
\centering
\includegraphics[width=6.8 in]{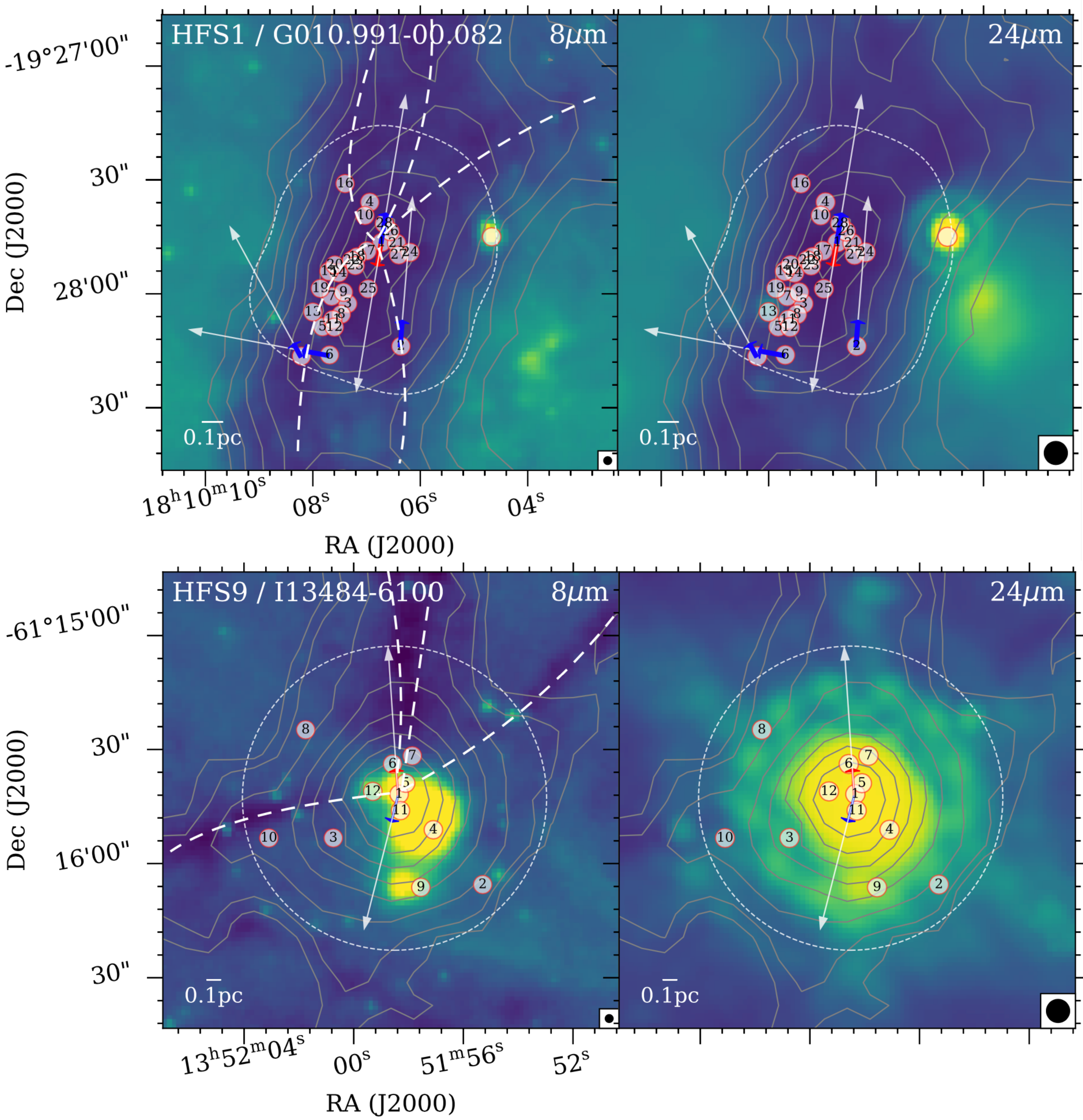}
\caption{Zoom-in images of Spitzer 8\,\um\ and 24\,\um\ for the central  regions of the IR-dark (in top row) and bright (in bottom row) HFS clouds.
The  contours represent 870\,\um\ dust continuum  from the ATLASGAL survey \citep{Sch09}. 
The blue/red arrows are blue/red-shifted outflowing lobes identified by \citet{Li 20}. 
They are purposely extended as white arrows for easy comparison of the relative orientation between the outflows and hub-composing filaments.
The dashed loop delineates the central subcloud field covered by our ALMA observations. The red circles indicate the cores identified from the ALMA continuum data. 
The dashed curves identify the filamentary structures.
The 8.0 and 24\,\um\ beams are shown at the bottom right-hand corner of the corresponding panel.
}
\label{fig:zoomin_hfs_outflows}
\end{figure*}

\begin{table}
\centering
\caption{Statistical number of outflows in HFSs.}
\label{tab:outflows}
\resizebox{8cm}{!}{
\begin{tabular}{cccc}
\hline\hline
HFS ID &  N (outflow lobes)  & Filament-aligned    & Filament-not-aligned (\%) \\
\hline
 HFS1 &   5  &   2  &   3 (60 )   \\
 HFS2 &  10  &   0  &  10 (100)   \\
 HFS3 &   3  &   0  &   3 (100)   \\
 HFS4 &   7  &   1  &   6 (86 )   \\
 HFS5 &  --  &  --  &  -- (-- )   \\
 HFS6 &   2  &   0  &   2 (100)   \\
 HFS7 &  13  &   1  &  12 (92 )   \\
 HFS8 &   3  &   0  &   3 (100)   \\
 HFS9 &   2  &   0  &   2 (100)   \\
HFS10 &   2  &   0  &   2 (100)   \\
HFS11 &   1  &   0  &   1 (100)   \\
HFS12 &   2  &   0  &   2 (100)   \\
HFS13 &   2  &   0  &   2 (100)   \\
HFS14 &   2  &   0  &   2 (100)   \\
HFS15 &   2  &   1  &   1 (50 )   \\
HFS16 &   2  &   0  &   2 (100)   \\
HFS17 &   7  &   1  &   6 (80 )   \\
\hline
\end{tabular}
}

\begin{flushleft}
{\bf Note:} The statistics was made from \citet{Li 20} and Baug et al. 2022 (under preparation).
\end{flushleft}
\end{table}

\subsection{Effects of outflow feedback on star formation}
\label{sec:outflows}
The final stellar mass depends not only on the initial mass reservoir of the natal clump but also on the mass accretion from the hub-composing filaments. However, several observational studies \egcite{Sch20} and theoretical simulations have shown the profound influence of stellar and protostellar feedback processes like  collimated jets and bipolar outflows \egcite{Wan10, Off17, Gus20, Gus21, Ver22}, and radiative heating \egcite{Bat09, Bat12, Kru12, Hen20, Hen22, Gru22} in inhibiting mass accretion.
Hence, for the protostar to grow in mass requires the strong accretion inflow to be least impacted by the above feedback processes in the early stages. As discussed in \citet{Dal11,Kum20}, ionizing radiation, stellar wind and ionizing gas (\hii\ regions) are found to channel out through pre-existing, inter-filamentary voids without dispelling the natal clump or inhibiting the mass inflow through filaments. 
In this study, we focus on the influence of outflows since these are pronounced in our HFSs sample.

The protostar's spin, and hence the orientation of the outflows, is inherited from the core scale where the angular momentum is hierarchically transferred from the natal cloud through the filament onto the star-forming core. Given that the large-scale filamentary inflow is either onto the short axes of the filament or along the long axis, the alignment of the outflows is expected to be either preferentially parallel or perpendicular to the filament.  \citet{Ana08} found observational evidence that outflows are preferentially aligned perpendicular to the filaments. In comparison, observational results presented by \citet{Dav09} and more recently by \citet{Ste17,Bau20} reveal no preferred outflow-filament orientation. Several other studies have shown that the rotation of the protostar could be independent of the parent filament \egcite{Tat16} or could evolve significantly during formation \egcite{Lee16,Off16}. The outflow-filament alignment has implication on the forming protocluster in HFSs. If the outflows run along the individual filaments, the filament-rooted longitudinal mass flows will be inhibited or halted which will significantly hinder the mass growth of young stars embedded in the central hubs \egcite{Wan10}. 

To understand the potential effect of outflows on star formation in the HFSs studied here, we compiled the parameters (i.e., position, and orientation)
of the associated outflows 
from \citet{Li 20} and Baug et al. (2022; under preparation). 
\citet{Li 20} have catalogued the CO and SiO outflows of the 8 IR-dark HFSs (i.e., HFSs\,1--8) 
taken from the ASHES survey data at $\sim 1.3$\,mm (see Sect.\,\ref{sec:observe}), while Baug et al. (2022) provide the \hcop\ outflows of the 9 IR-bright HFSs (i.e., HFSs\,9--17) 
taken from the ATOMS survey data at $\sim 3$\,mm (see Sect.\,\ref{sec:observe}).
As indicated in Table\,\ref{tab:outflows}, we obtain 60 outflow lobes for all the HFSs investigated here, where 38 and 22 outflow lobes are associated with the IR-dark and IR-bright HFSs, respectively. Note that we have considered the individual lobes of outflows here instead of the entire outflow entities. 

The location and orientation of the identified outflows are displayed in Fig.\,\ref{fig:zoomin_hfs_outflows}, where the red/blue arrows indicate the estimated orientation and extent of the red/blue-shifted lobes. 
As evident from the figure, most of the outflows have spatial extents  smaller than the dimension of the host HFS subclouds (i.e., the central dense regions covered by ALMA observations). 
This result indicates that the outflows  might have not  escaped the dense regions of the HFSs in early stages of star formation.
However, this result may be an observational consequence due to the lack of
total power in the analysis made in the ASHES and ATOMS surveys. 
 But, if the outflows can escape the dense regions of the HFSs, then their direction can be traced by extending the identified outflows. These are shown as white arrows in the figure.

The statistics of the outflows and their orientation are tabulated in Table\,\ref{tab:outflows}.
For HFS\,5 there are no identifiable outflows while
HFS\,1 and 15 have 60\,\%, and 50\% of filament-aligned outflows, respectively. Except for these three HFSs, all the other HFSs have the majority of outflows (i.e., $\ge86$\%)  not aligned with the individual filaments.  There are a couple of caveats in the above analysis which need to be highlighted here. Firstly, the outflow identifications in the  two studies used here are probably incomplete due to the presence of multiple overlapping outflows which tend to occur in high-mass star formation regions. Secondly, projection effects would also influence the observed orientations. Last but not the least, one needs to study more number of HFSs-outflow systems to enable a more robust statistical investigation. Keeping the above caveats in mind, our results show that the observed outflows (1) do not preferentially align parallel or perpendicular to the filaments and (2) tend to be oriented toward the voids of the hub-composing filaments. This suggests that, similar to the effect of other feedback processes, outflows render a limited effect on filamentary mass inflow and thus on the mass growth of young stars being formed in the centrally located hubs. 

Moreover, the inference obtained above agrees with the quantitative analysis of the related energies. 
For the IR-dark HFSs, \citet{Li 20} found that the outflow-induced turbulence cannot sustain the internal turbulence of the natal clumps 
as the outflow energy rate is around two orders of magnitude less than the turbulent energy dissipation rate. These authors also infer the outflow energy to be much smaller than the gravitational energy of the clumps.
For the more evolved IR-bright HFSs investigated here, three of which (i.e.., HFSs\,11--13) were studied in terms of the outflow dynamics in \citet{Bau20}, they argued that 
the kinetic energy of outflows alone cannot balance the gravitational binding energy of the hosting clumps.
Taken together, these results indicate the limited effect of outflows on the destruction of their host HFSs in early stages and thus on the progress of star formation therein.  However, as mentioned above, a larger sample and improved statistics on filament-outflow alignment is required to conclusively interpret simulations of outflow feedback \egcite{Wan10, Off17, Gus20, Gus21, Ver22} in the context of HFSs.

\section{Discussion}
\subsection{Spatial distribution of cores and YSOs}
\label{sec:core_spatial_distr}

\begin{figure}
\centering
\includegraphics[width=3.4 in]{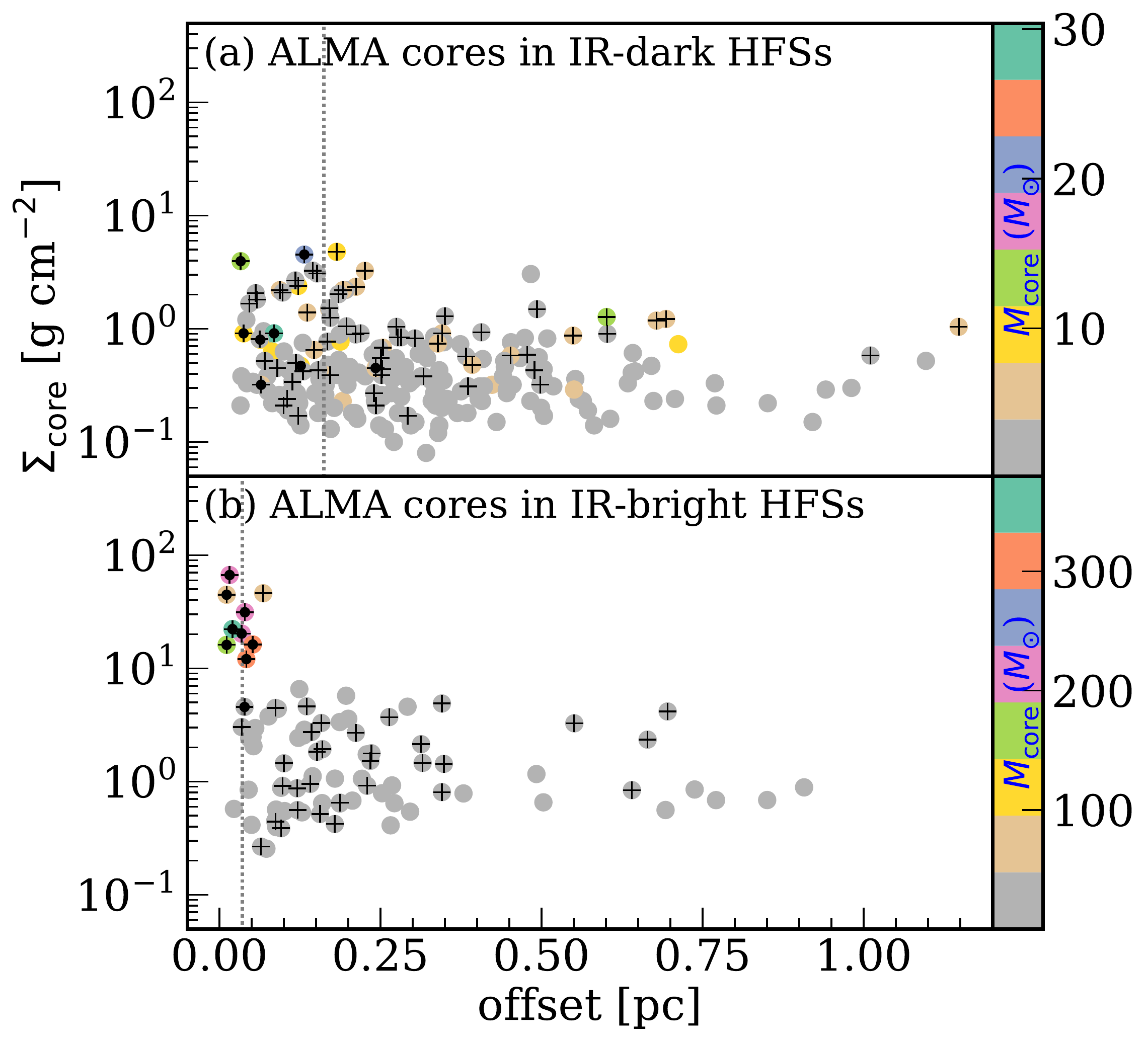}
\caption{ Distribution of mass surface density of cores in the HFSs
against distance from the HFS centre. The colors in circles reflect the mass distribution of the cores. Panels\,(a) and (b) display  the cores in
the IR-dark, and IR-bright HFSs, respectively. 
In both panels, the protostellar cores are indicated in circles with inserted pluses while candidate starless cores are in empty circles. Circles with black dots inside are the centrally located most massive cores.
The vertical dotted lines indicate the average distance weighted by the mass surface density of cores,
i.e., 0.16\,pc in panel\,a, and 0.04\,pc in panel\,b.
}
\label{fig:core_M_Sig}
\end{figure}

\begin{figure*}
\centering
\includegraphics[width=6.8 in]{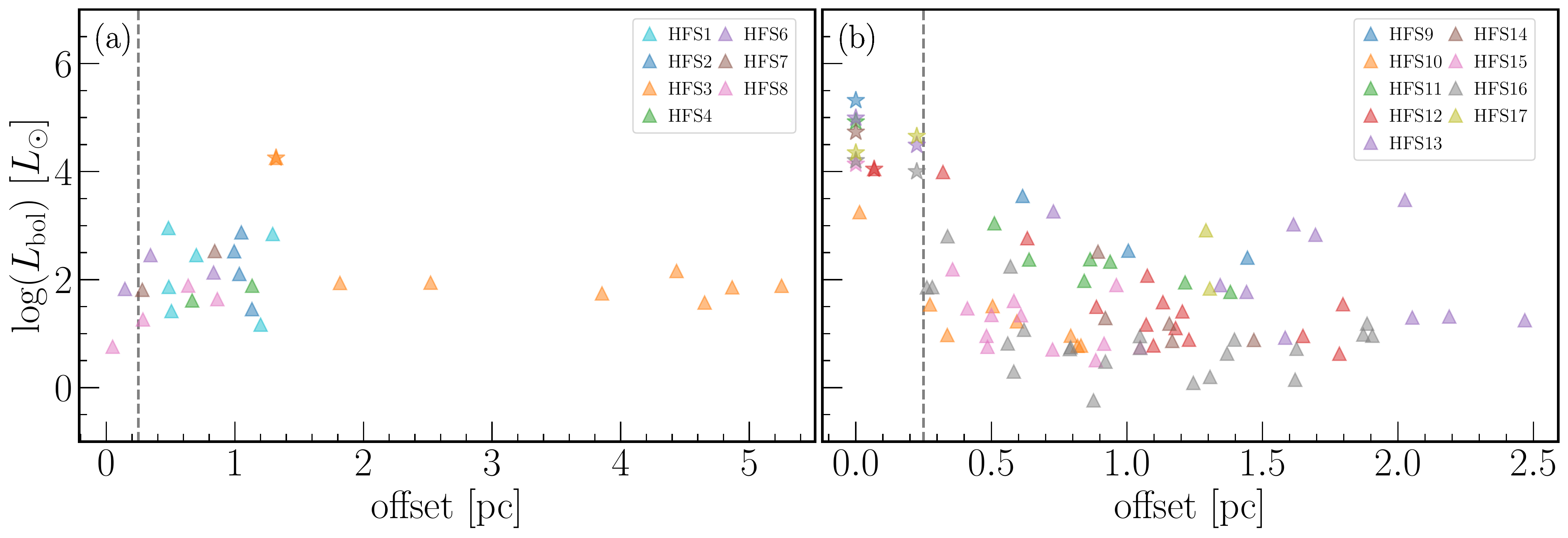}
\caption{Luminosity of YSOs bright at 24\,\um\ associated with the IR-dark (panel\,a) and IR-bright HFSs (panel\,b) as a function 
of the distance from the HFS centre. The dashed lines indicate the typical size of the centrally-located clump for all HFSs, as defined in Fig.\,\ref{fig:core_M_Sig}.  
 Stars having luminosity above $10^4$\,\lsun\ are shown  as star symbols 
as opposed to  those having luminosity below  $10^4$\,\lsun\  shown as triangle symbols.
}
\label{fig:yso_distr}
\end{figure*}

\vspace{1mm}
\noindent \textbf{Cores in the HFSs} \par
 \noindent Figure\,\ref{fig:core_M_Sig} shows the distribution of the mass surface density ($\Sigma_{\rm core}$) of the cores as a function of the distance from the centre of the host HFSs. The centre is defined to be the position where the intensity of the 870\,\um\ emission peaks. Circles with inserted plus symbols distinguish the protostellar cores from that of the candidate starless ones. The centrally located most massive core of each HFS has an additional dot symbol included. Further, the plotted circles are colour-coded to represent the core mass ($M_{\rm core}$) distribution. Several interesting trends can be deciphered from these plots and are discussed below. 

The number of cores is more in the central region with only a sparse population seen beyond $\sim$0.5~pc. The massive and dense cores in the IR-dark HFSs are located within $\sim$0.25~pc, whereas, in IR-bright HFSs these are confined to the innermost region of $\sim$0.05~pc. All nine centrally located most massive cores in IR-bright HFSs are forming high-mass stars as inferred from the associated high luminosities of $>10^4$\,\lsun. This supports the scenario that in HFSs, the central areas of hubs are preferential sites for high-mass star formation where mass accretion occurs from the hub-composing filaments. The ideal location of such cores in IR-dark HFSs also qualifies them as potential high-mass star-forming cores.

The steep gradient seen in the spatial distribution of the most massive and dense cores towards the inner most region in IR-bright HFSs suggests a more centrally peaked clustering as opposed to the wider distribution observed in the IR-dark HFSs. That is, the spatial distribution of massive dense cores peaks at $\sim 0.16$\,pc in IR-dark HFSs, but at $\sim 0.04$\,pc in IR-bright HFSs, as indicated in the dotted lines in the figure. These represent the average distance from the HFS centre weighted by the mass surface density over all the cores in each IR type of HFSs. The wider distribution of massive cores in the IR-dark hubs is in good agreement with the results of \citet{San19}. These authors propose that cores in IR-dark HFSs originate from hierarchical subclustering rather than from centrally peaked clustering. The observed difference in the two IR types could suggest transformation to a centrally peaked clustering following the evolution of the host HFSs from the IR-dark to IR-bright stages.

 Scarcity of high-mass prestellar cores (of $M_{\rm core}\ge30$\,\msun\ over the 0.1\,pc scale, e.g., \citealt{San17,San19}), the progenitors of high-mass stars, is observed in our sample of IR-dark HFSs. Here, none of the detected cores have masses greater than the threshold defined above. This allows us to conjecture that high-mass star formation could involve a dynamical, continuous mass accretion with evolution, which will be discussed further in Sect.\,\ref{sec:mass_transfer}. If we consider the prestellar cores in the IR-bright sample, only 11 ($\sim 13\%$ of cores) have mass estimates greater than $30$\,\msun. For these to form high-mass stars, the same mass accretion process should ensue. 
 However, the starless or prestellar nature of those cores needs to be confirmed through future higher-resolution observations with a more sensitive outflow tracer (e.g., CO~1--0),


\vspace{1mm}
\noindent \textbf{YSOs in the HFS clouds} \par

\noindent Figure\,\ref{fig:yso_distr} presents the distribution of bolometric luminosity ($L_{\rm bol}$) of candidate YSOs against the distance from the centre
of the host HFSs. 
In the IR-dark HFSs, the YSO's luminosity distribution is nearly constant at a low luminosity level (i.e., $\sim 100$\,\lsun) typical of low-mass protostars, regardless of the distance of the YSOs from the centre of the HFSs. Only one YSO with $L_{\rm bol}\sim10^4$\,\lsun, that is typical of high-mass protostars, is found at distance $>1$\,pc from the
centre of the host HFSs.
Furthermore, only two HFSs (i.e., HFSs\,6, and 8) have identified YSOs (one each)
within the hub-clump (i.e., central clump in the hub) region 
and  their luminosities are low (i.e., $L_{\rm bol}\la 100$\,\lsun).
In  the case of the IR-bright HFSs, except for the three YSOs having $L_{\rm bol}$  $\sim 10^3$--$10^4$\,\lsun\ located in a distance range of [1.6, 2.1]\,pc from the center of HFS\,13,
all of the YSOs show a decreasing trend in luminosity from high ($\ga10^4$\,\lsun) to low ($<100$\,\lsun) luminosity values
with the distance from the centre of the host HFSs up to 
$\sim 2$\,pc, beyond which the YSOs display nearly constant low luminosity values. Note that given the not so high luminsities (i.e., $<10^4$\,\lsun) of the above mentioned three luminous YSOs 
far from the centre of the host HFS\,13, we assume that they could represent a cluster of intermediate and/or low-mass young stars instead of high-mass protostars, which agrees with the apparent multiplicity of these three YSOs seen at 8\,\um\ but not well resolved at 24\,\um. In view of this, the observed decreasing trend suggests a luminosity/mass-segregated cluster formation picture
in the IR-bright stage of HFSs, in which high-mass stars represented by high luminosities prefer to form in the central area of HFSs (i.e., the hub-clump region), while low-mass stars represented by low luminosities tend to form in the outskirts of HFSs up to several pc. In addition, the number of high-luminosity YSOs of $L_{\rm bol}>10^4$\,\lsun\ found in IR-bright HFSs is larger (eleven; see star symbols in Fig.\,\ref{fig:yso_distr}b) compared to the IR-dark HFSs where only one are  detected.
Moreover, almost each of the IR-bright HFSs has a corresponding high-luminosity YSO within the hub-clump region, in contrast to the absence of high-luminosity
YSOs within the same region of the IR-dark HFSs. The above distribution of YSOs possibly implies an evolutionary sequence from a relatively quiescent, IR-dark phase to an active, IR-bright phase.

\subsection{Multi-scale mass transfer and high-mass star formation}
\label{sec:mass_transfer}

\begin{figure*}
\centering
\includegraphics[width=3.4 in]{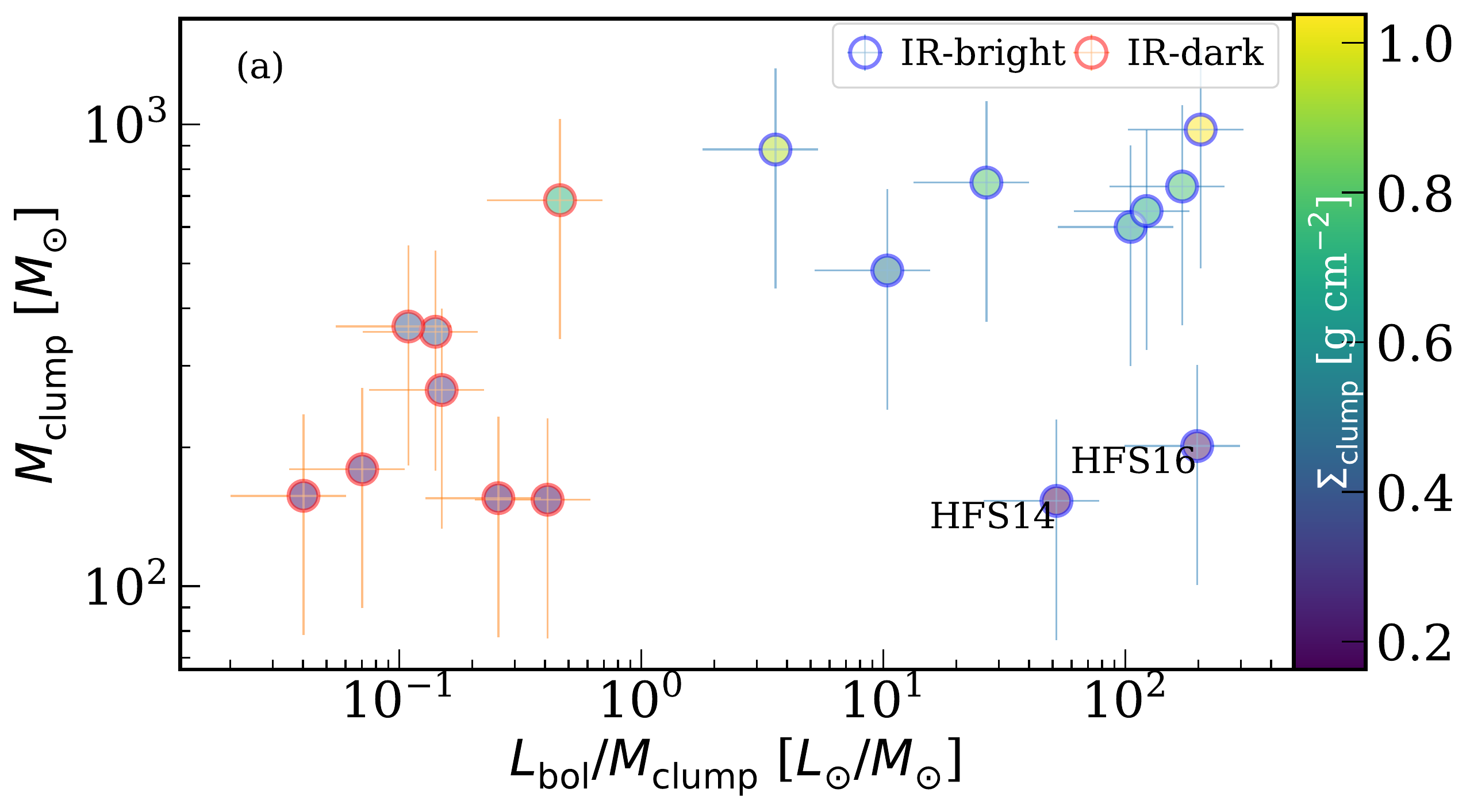}
\includegraphics[width=3.4 in]{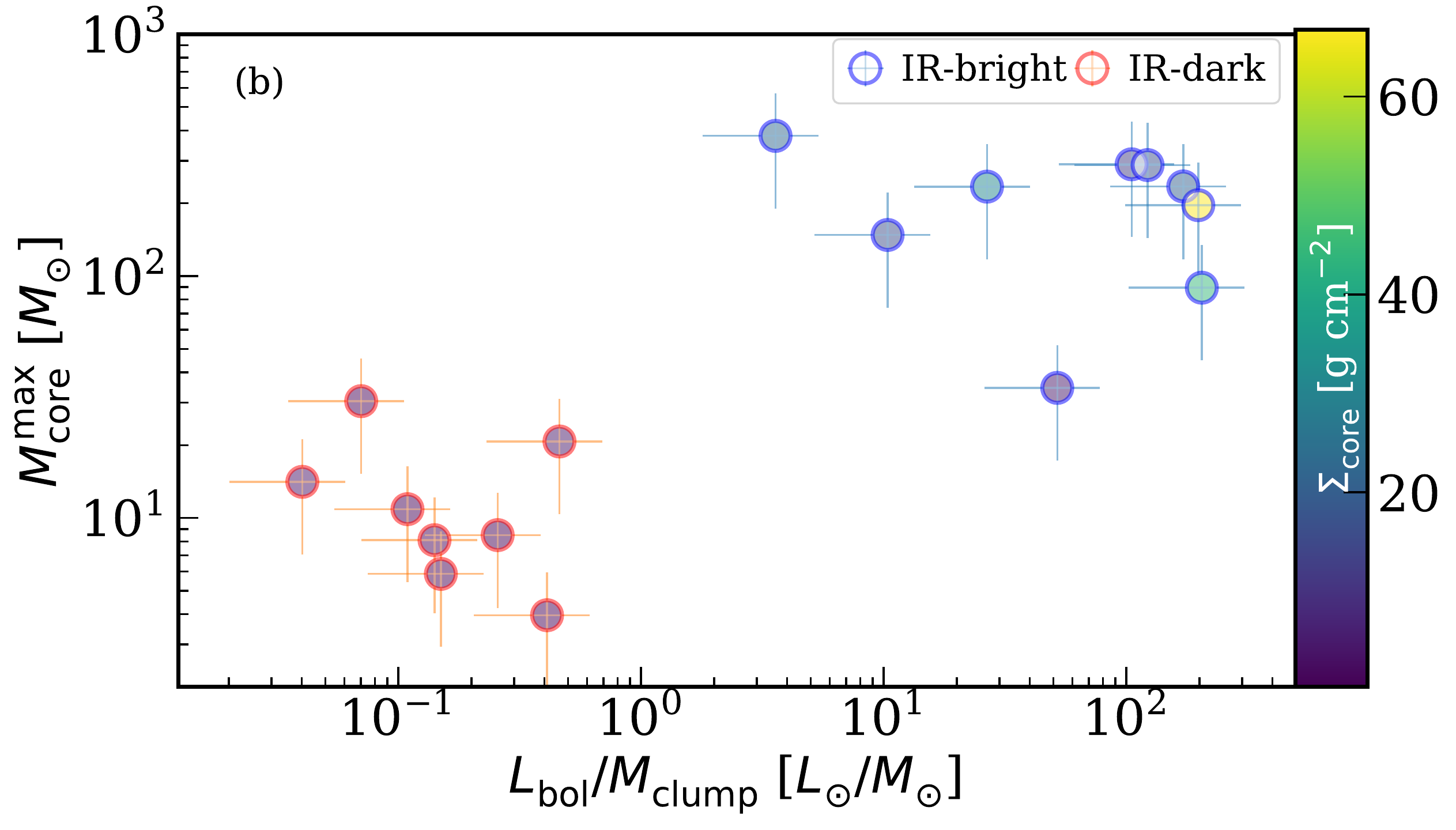}
\caption{Mass distribution of the central massive clumps (panel\,a) and their embedded most massive cores (panel\,b) in the two IR types of HFS against the 
the $L_{\rm bol}/M_{\rm clump}$ ratio of the clumps. The color-coded circles  reflect the mass surface density of the sources. 
The $L_{\rm bol}/M_{\rm clump}$ ratio of the clumps represent  the evolutionary stage of high-mass star formation of the HFSs.
}
\label{fig:mass_correlate}
\end{figure*}
Figure\,\ref{fig:mass_correlate} presents the mass distribution of the central massive clumps of the HFSs (in panel\,a) and their embedded massive cores (in panel\,b)
against the $L_{\rm bol}/M_{\rm clump}$ ratio of the clumps.
The colors in circles in the figure represent the mass surface density of the sources. Note that, for this analysis, we have only considered the most massive cores
since 1) they are (potential) sites of high-mass star formation, and 2) their mass and surface density estimates are not significantly affected by the observation 
bias caused by the different MRSs associated with the ASHES and ATOMS data (see Sect.\,\ref{sec:core_params_derived}).
As shown in the figure, the IR-dark and IR-bright stages of HFSs can be well represented by the $L_{\rm bol}/M_{\rm clump}$ ratios of the clumps. 
The mass and mass surface density of the central clumps show a marginal increase of a factor of $\sim 3$ (on average) from the IR-dark to IR-bright stage. In comparison, for the embedded massive cores, the estimated values of the above parameters are enhanced
by a factor of $\sim 24$ (on average) from the IR-dark to IR-bright stage. 
Additionally, as previously discussed (see Sect.\,\ref{sec:core_spatial_distr}), high-mass protostars (i.e., with  $L_{\rm bol}\ga10^{4}$\,\lsun) are only identified in the IR-bright HFSs.
These results suggest that sufficient mass accumulation from the large-scale, hub-composing filaments is required for the central clumps in IR-dark HFSs to evolve to clumps with high-mass protostars in IR-bright HFSs. This process would continue till the hub-composing filaments are completely dispersed
by stellar feedback (e.g., stellar winds, and ionization).
Further, the associated massive cores accumulate the required
mass from their natal clumps. 
Thus, a multi-scale mass accretion/transfer scenario
unfolds in HFSs, where the mass accretion/transfer proceeds from the large-scale hub-composing filaments, through clumps, down to 
cores where high-mass stars finally form. 
Consistent with the observed trend, the mass and mass surface density of the clumps, and cores should be higher
in the IR-bright stage of HFSs 
since the accretion  timescales of these density structures 
are more extended in the more evolved stage as along as the large-scale hub-composing filaments contain sufficient gas material.
This multi-scale mass accretion process has also been observed toward one of
the HFSs studied here (i.e., HFS\,17) in \citet{Liu22a}. These authors reveal the presence 
of the multi-scale mass accretion flows, i.e., accretion from clumps onto cores, and that from cores to embedded protostars.

The above results from  a selected sub-sample of the ASHES and ATOMS surveys agree well with the filament to cluster (i.e., F2C) evolutionary sequence discussed in a recent statistical study by \citet{Kum20}. Based on a large sample of $\sim 3700$ candidate HFSs using far-infrared {\it Herschel} dust continuum maps at 70--500\,\um\ from the Hi-Gal survey, these authors propose four stages involved in the formation of high-mass stars in the context of HFSs. These are:
I) formation of individual dense filaments
by mechanisms such as cloud-cloud collisions, and compression from local turbulence; II) flow driven filaments overlap wherein intra-filamentary matter in the HFS cloud combine to form a hub with density amplification making them more conducive to star formation as compared to the filaments;
III) formation of high-mass stars in the density amplified hub where the generated gravitational potential difference between the hub and the filaments
can trigger and direct the filament-rooted longitudinal flows toward the centrally-located hub.
IV) formation of ``classical'' (optically visible) \hii\ regions in the hub along with a small embedded cluster of stars. 
In this stage, the radiation pressure and ionization feedback from the newly forming massive stars channel out of the hub through the inter-filamentary diffuse cavities.
These four stages, where a multi-scale mass accretion/transfer process can be expected from hub-composing filaments through clumps (hubs) to cores (i.e., Stage\,II and III),
finally lead to
a mass-segregated embedded cluster with high-mass stars preferentially formed in the hub and low-mass stars in the
hub-composing filaments. 

From the observational study presented here, the IR-dark HFSs resemble Stage\,II, where the density-enhanced hub has formed and is in a relatively quiescent phase of star formation. Presence of low-luminosity YSOs outside the hub-clump region implies onset of low-mass star formation in the HFS cloud and/or individual filaments while the longitudinal flows continue to feed matter to the central hub which are devoid of high-luminosity sources. In comparison, the observational features seen in IR-bright HFSs are characteristic of Stage\,III. In this sample, in addition to a similar picture of low-mass star formation in the entire HFS cloud, a small, mass-segregated embedded cluster of YSOs (see Fig.\,\ref{fig:yso_distr}b), in which high-luminosity YSOs ($\ga10^4$\,\lsun) typical of high-mass protostars are preferentially located in the hub-clump region. Interestingly, the orientation of outflows along the low density, inter-filamentary voids (see Sect.\,\ref{sec:outflows}) also gives clues for channelling out radiation pressure and ionization feedback in the next evolutionary stage (IV) of the classical \hii\ regions.

 Recent studies of molecular clouds found evidence for multi-scale hierarchical fragmentation cascade (i.e., from clouds, through filaments, clumps and cores, down to protostars, see e.g., \citealt{Eli18, Tho22}) probably as a major vector of star formation. In conjunction with the latest
 theoretical models such as GHC \citep{Vaz19} and I2 \citep{Pad20},  there seems to be a general consensus which can  favor high-mass star formation in HFSs through a multi-scale mass accretion/transfer process that finally can lead to a mass-segregated cluster of stars.  Notwithstanding the selection bias (see Sect.\,\ref{sec:core_params_derived}), one may consider the observed distinct mass distribution of the central massive clumps and their most massive cores as an evidence for the above processes in play in HFSs.
  Towards the above efforts and to put more robust constraints to theoretical models, companion papers (e.g., Yang, D.T. et al. 2022, in prep.) are in the pipeline on high-resolution, multi-scale (i.e., from hub-composing filaments, clumps, to cores) kinematic and dynamical studies dedicated to the HFSs investigated here using the spectral line data from the same surveys.
   For example, the GHC and I2 models  agree on gravity-driven mass--accretion on small scales (e.g., cores), however, they predict two distinct drivers on larger scales for the multi-scale mass accretion process. The former strongly favors a gravity--driven hierarchical mass accretion
    while the latter advocates for a turbulence--driven mass inflow/accretion, which can be disentangled with the multi-scale kinematic and dynamical studies.

\section{Summary and conclusions} \label{sec:summary}
We have presented a statistical study  of a sample of 17 high-mass star formation HFSs using high-angular resolution ($\sim 1$--2\arcsec) ALMA 1.3\,mm and 3\,mm continuum data.
The statistical results have helped shed light on the high-mass star formation scenario in HFSs. Our main results can be summarised as  follows:

\begin{itemize}

     \item The 17 HFSs are selected from the target lists of the ASHES 1.3\,mm and ATOMS 3\,mm surveys. They are  identifiable in the Spitzer 8\,\um\  image with hub-composing filaments intersecting at the central hub. All the hub-composing filaments appear  as elongated dark lanes in 8\,\um\ emission. Based on the different IR types of the hubs,
     the HFSs are divided into 8 IR-dark and 9  IR-bright HFSs. The IR-dark HFSs contain an IR-dark hub  without detectable IR emission shortward of 70\,\um,
     while the IR-bright  HFSs have an IR-bright hub  with  high-mass protostars in the same wavelength regime. The two IR types can represent an evolutionary sequence of high-mass star formation HFSs from the IR-dark to IR-bright  stage. 
     \item The 17 central massive clumps are identified in their natal HFSs
      from the available ATLASGAL 870\,\um\ continuum data. In addition, 310
      embedded cores are extracted from the ALMA  continuum data, including 224
      from the IR-dark HFSs, and 86 from the IR-bright HFSs.

     \item The massive, dense cores in the two IR types of HFSs are predominantly
     distributed in the central hub-clump region of HFSs of radius 0.25\,pc. For IR-dark HFSs, the cores peak within $\sim 0.16$\,pc of the centre displaying a hierarchical sub-clustering mode. This transforms to a centrally-peaked clustering mode in IR-bright HFSs where the cores peak within $\sim 0.04$\,pc of the centre.

    \item The central massive clumps and their associated most massive cores in HFSs  show a trend of  increasing mass and mass surface density with the evolution of 
    HFSs from the IR-dark to  IR-bright stage. This could be a natural result of
    the multi-scale mass accretion/transfer scenario in HFSs from the hub-composing filaments through clumps to cores. 

    \item A total of 122 candidate YSOs associated with the 17 HFSs are retrieved from the combined catalogues of the archival Spitzer/MIPSGAL 24\,\um\ point sources and the Spitzer/IRAC candidate YSOs. Their stellar bolometric luminosities are estimated from the 
    24\,\um\ flux.
    From the spatial distributions of YSOs in the HFSs, we find the picture of a mass-segregated cluster of YSOs in which high-luminosity YSOs typical of high-mass protostars are preferentially located in the central hub-clump region, and surrounded by a population of low-luminosity YSOs typical of low-mass protostars in the entire HFS cloud extending to several parsecs.

    \item From qualitative  analysis of the relative orientation between the outflow and hub-composing filaments in all the HFSs studied here, most of the outflows are found oriented toward the  lower density inter-filamentary cavities. This suggests that outflow feedback  would have a limited effect on the disruption of the HFS clouds and ongoing star formation therein.

\end{itemize}

 From the observed  facts of the trend on multi-scales (i.e., clumps and cores) of  increasing mass and mass surface density  with evolution from IR-dark to  IR-bright stage, the mass-segregated cluster of YSOs, and
the preferential escape directions of outflow feedback, we conclude that high-mass star formation in the HFSs can be described by a multi-scale mass accretion/transfer scenario, from hub-composing filaments through clumps down to cores, that can naturally lead to a mass-segregated cluster of stars. To reveal the detailed physics related to the multi-scale accretion scenario requires  further investigations, which will be carried out  in our future multi-scale kinematic and dynamical studies dedicated to the HFSs investigated here using the high-resolution spectral line data from both ATOMS and ASHES
surveys.

\medskip
\noindent{\textbf{ACKNOWLEDGEMENTS}}\\
 We thank the anonymous referee for comments and suggestions that greatly improved the quality of this paper.
This work has been supported by the National Key R\&D Program of China (No. 2022YFA1603101).
H.-L. Liu is supported by National Natural Science Foundation of China (NSFC) through the grant No.12103045.
T. Liu acknowledges the supports by NSFC through grants No.12073061 and No.12122307.
 PS was partially supported by a Grant-in-Aid for Scientific Research (KAKENHI Number 22H01271) of the Japan Society for the Promotion of Science (JSPS).
S.-L. Qin is supported by NSFC under No.12033005.
AS gratefully acknowledges support by the Fondecyt Regular (project
code 1220610).
This research was carried out in part at the Jet Propulsion Laboratory, which is operated by the California Institute of Technology under a contract with the National Aeronautics and Space Administration
(80NM0018D0004).
G.G., AS and L.B. gratefully acknowledges support by the ANID BASAL projects ACE210002 and FB210003.
C.W.L. is supported by the Basic Science Research Program through the National Research Foundation of Korea (NRF) funded by the Ministry of Education, Science and Technology(NRF-2019R1A2C1010851), and by the Korea Astronomy and Space Science Institute grant funded by the Korea government (MSIT) (Project No. 2022-1-840-05).
This work is supported by the international partnership program of Chinese Academy of Sciences through grant No.114231KYSB20200009, and Shanghai Pujiang Program 20PJ1415500.
This paper makes use of the following ALMA data: ADS/JAO.ALMA\#2019.1.00685.S and 2015.1.01539.S. ALMA is a partnership of ESO (representing its member states), NSF (USA), 
and NINS (Japan), together with NRC (Canada), MOST and ASIAA (Taiwan), and KASI (Republic of Korea), in cooperation with the Republic of Chile. The Joint 
ALMA Observatory is operated by ESO, AUI/NRAO, and NAOJ.
This research made use of astrodendro, a Python package to compute dendrograms of Astronomical data ({\url{http://www.dendrograms.org/}}).
This research made use of Astropy,
a community-developed core Python package for Astronomy \citep{Ast18}. \par

\vspace{2mm}
\noindent{\textbf{DATA AVAILABILITY}}\\
The data underlying this article will be shared on reasonable request
to the corresponding author.

\vspace{-5mm}

\input bbl
\appendix
\section{Complementary figures}
\label{sec:app-kin-dist}

\begin{figure*}
\centering
\includegraphics[width=2.0 in]{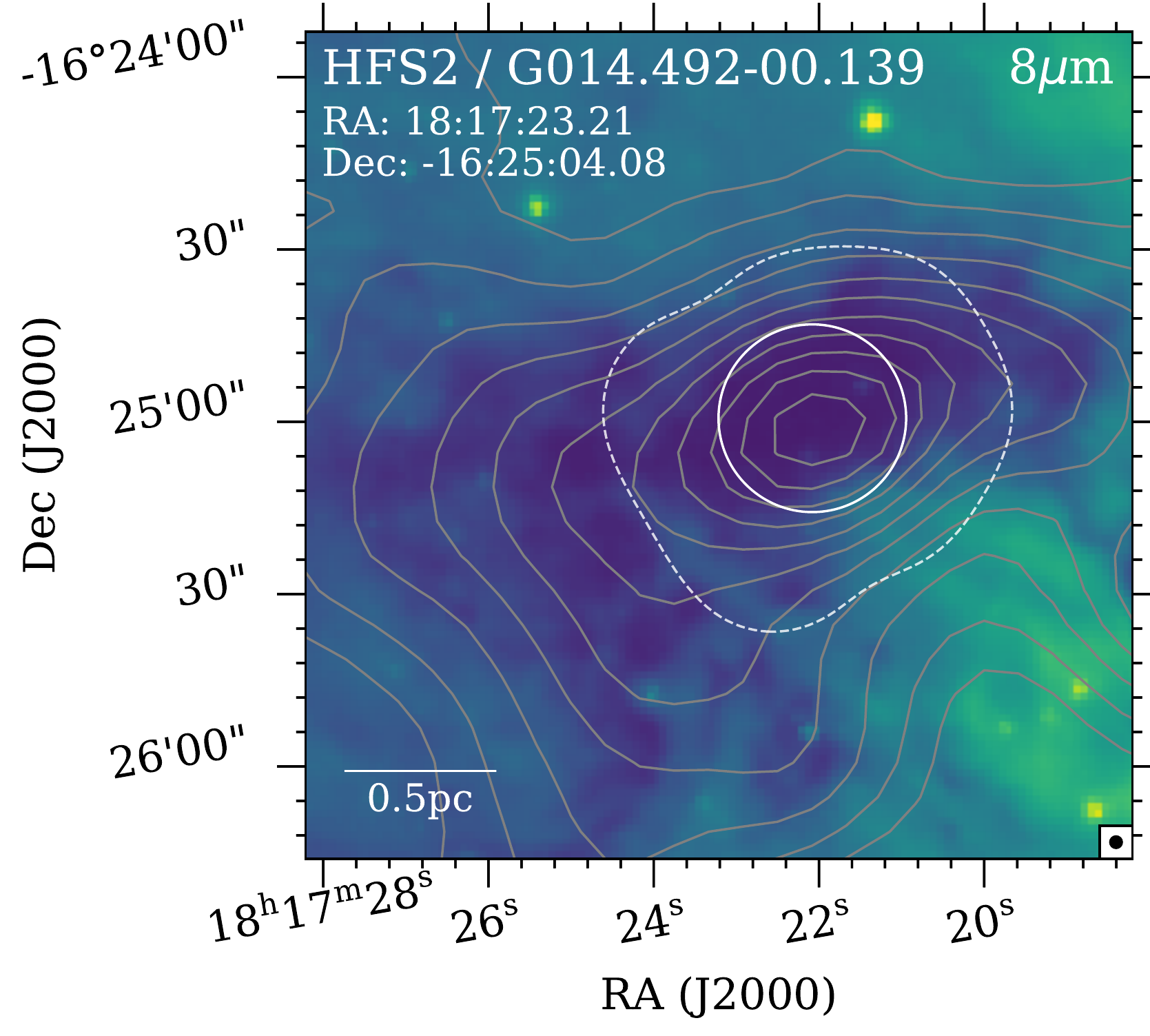}
\includegraphics[width=2.0 in]{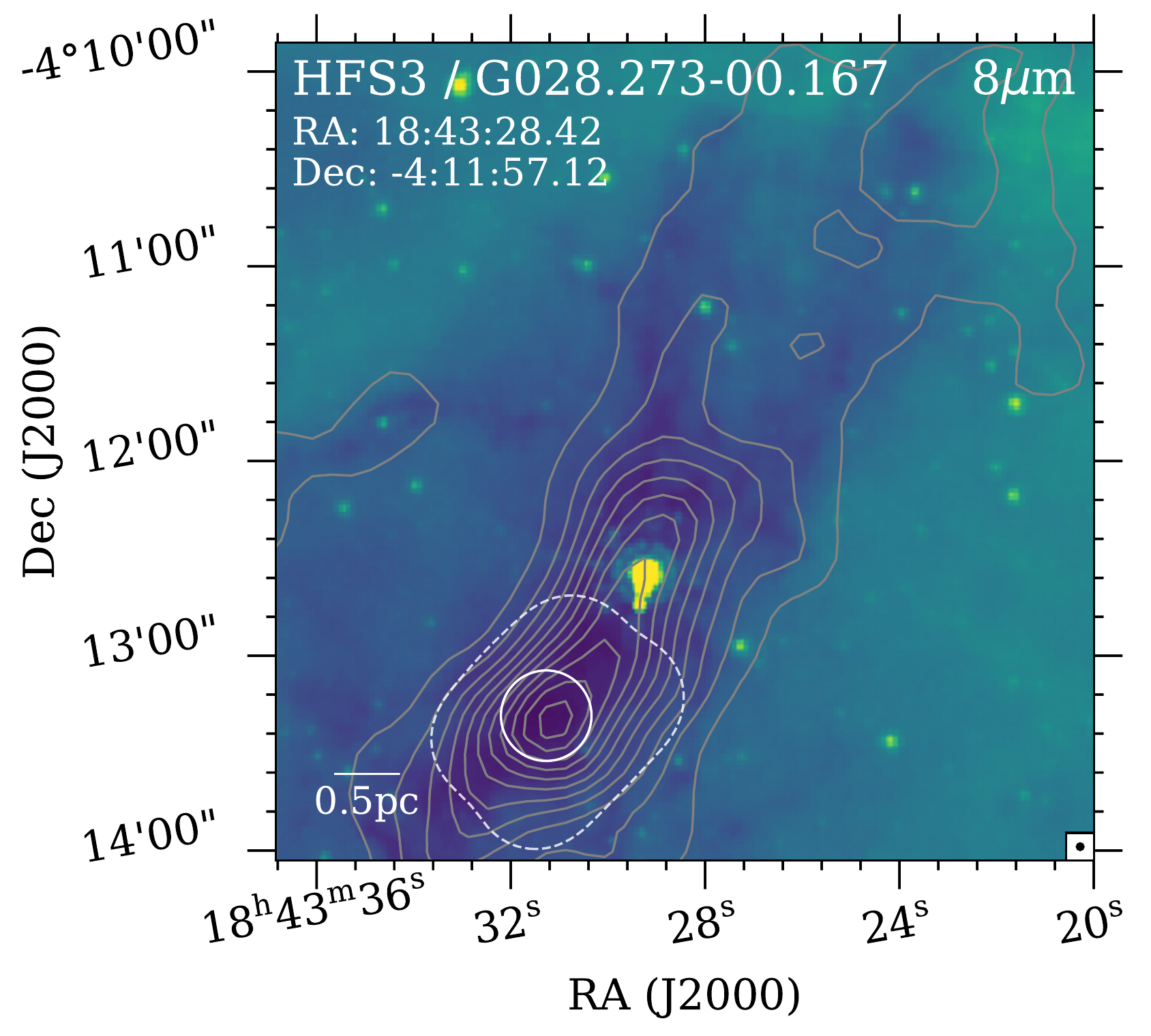}
\includegraphics[width=2.0 in]{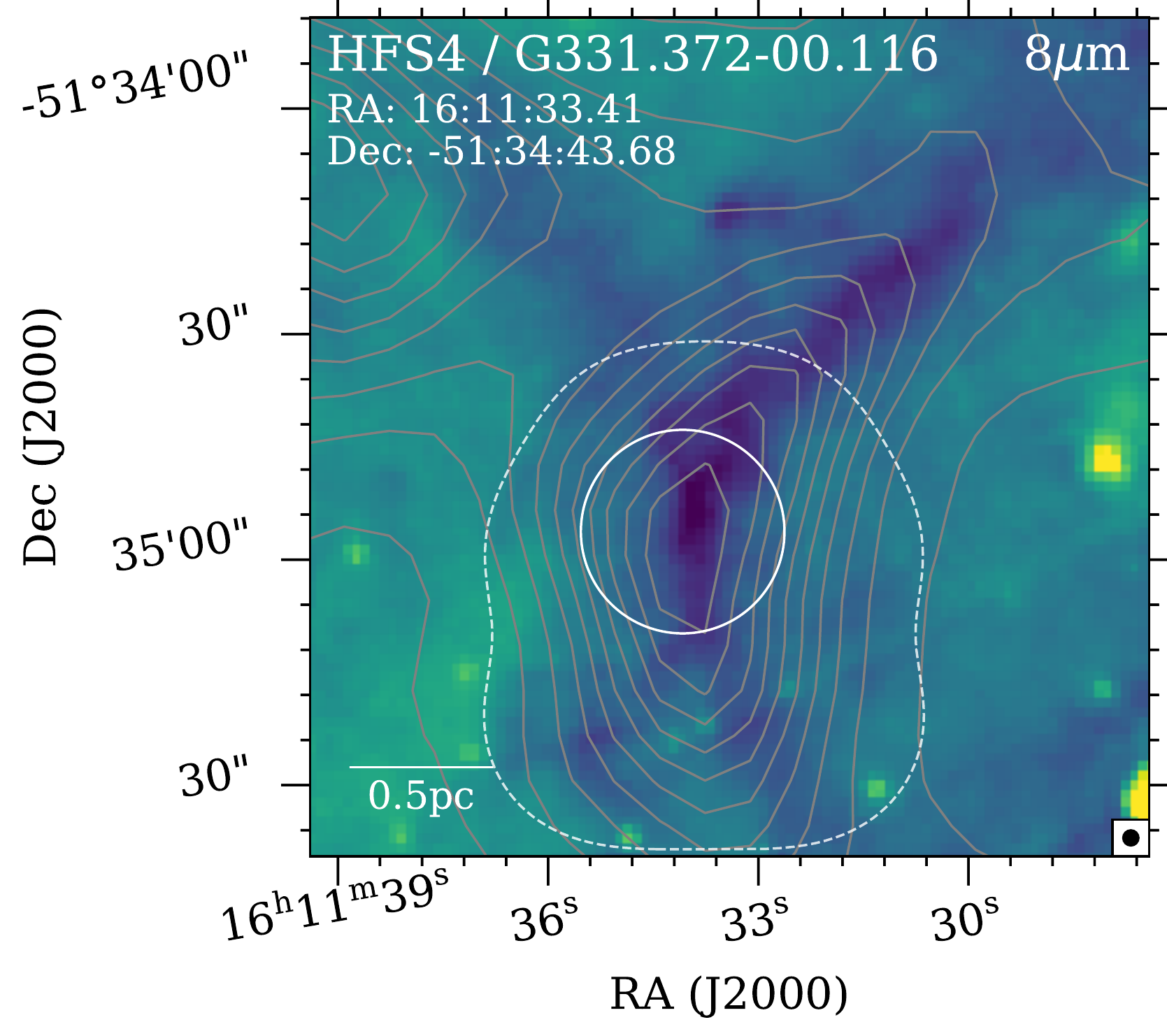}
\includegraphics[width=2.0 in]{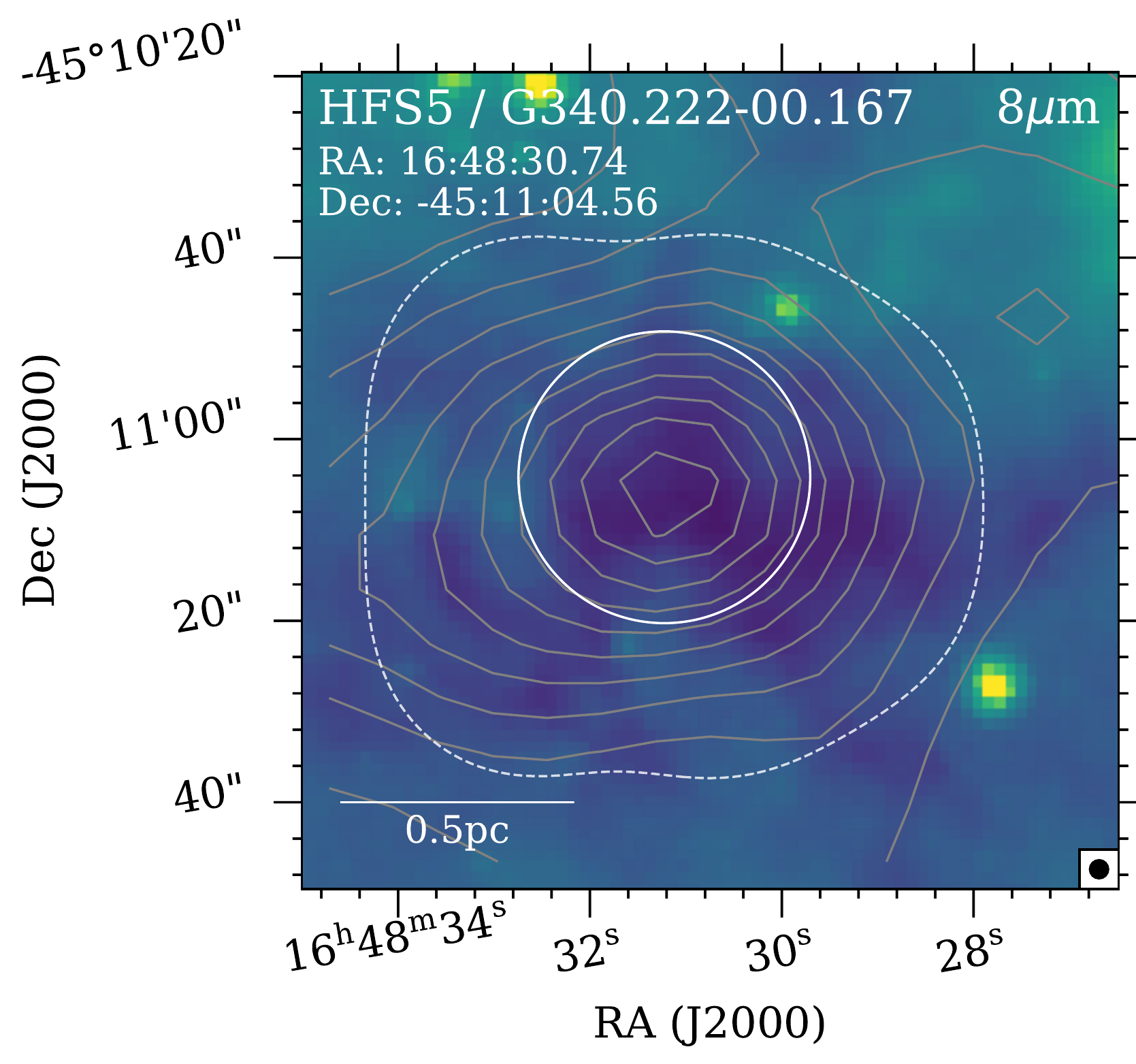}
\includegraphics[width=2.0 in]{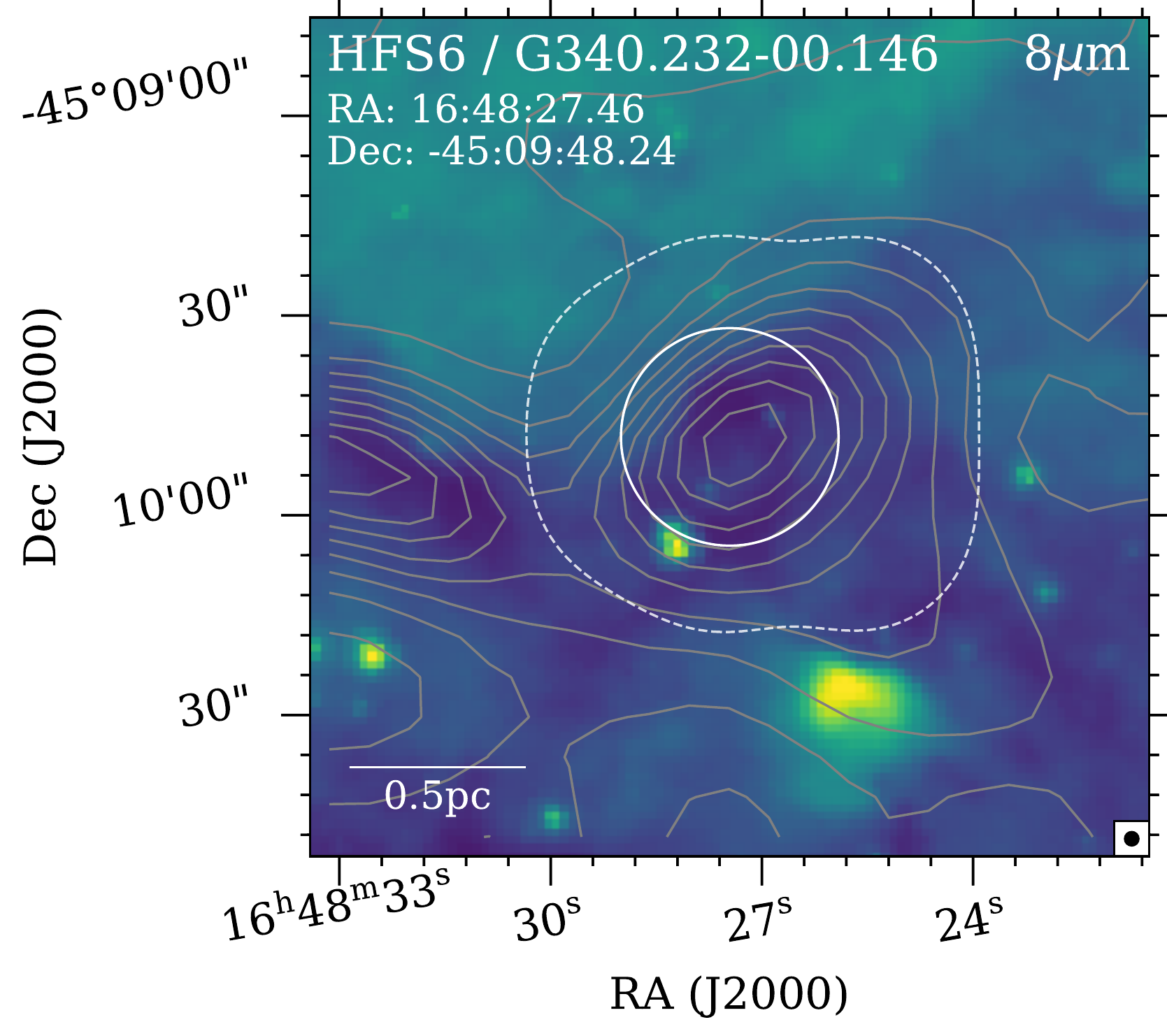}
\includegraphics[width=2.0 in]{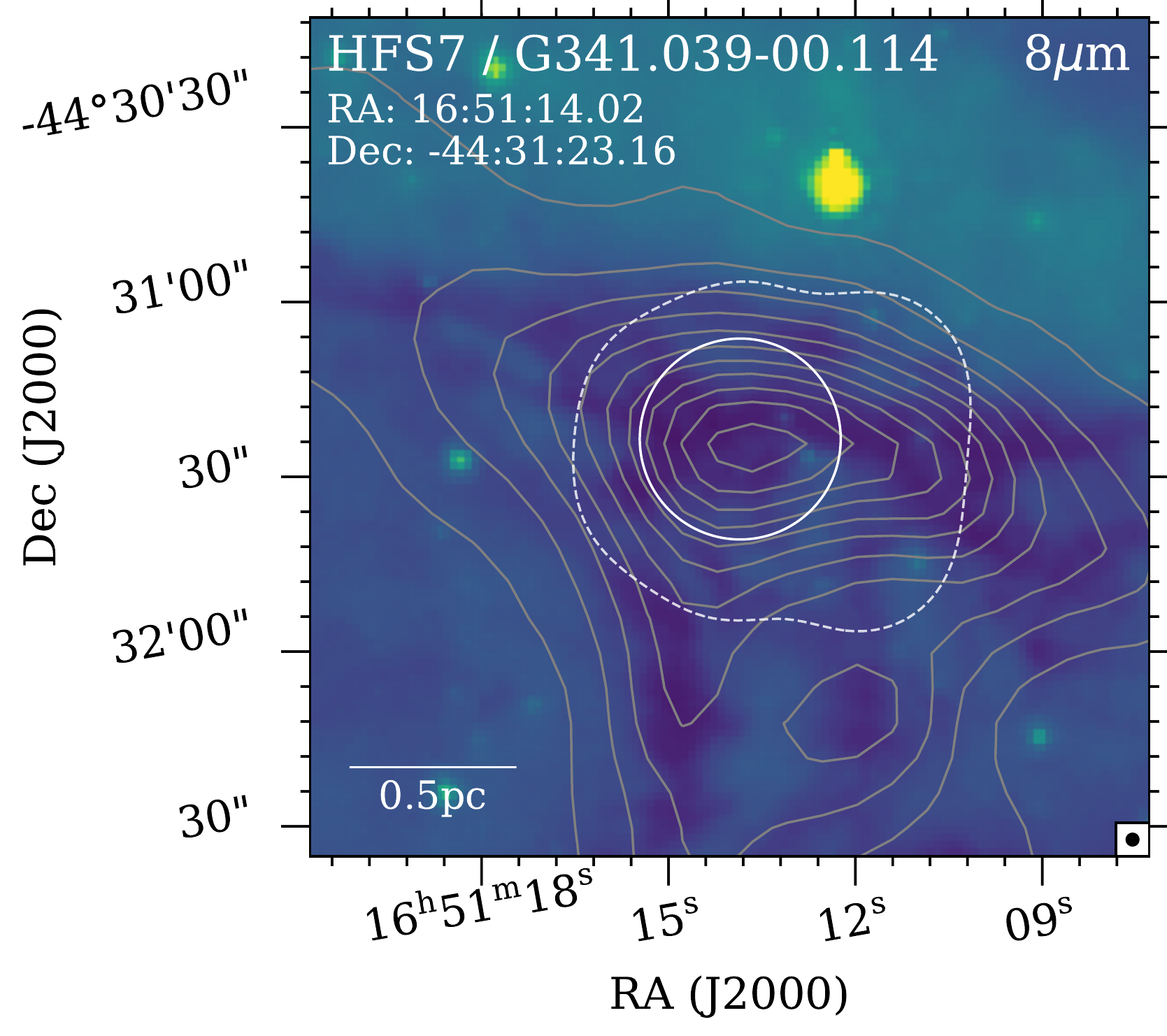}
\includegraphics[width=2.0 in]{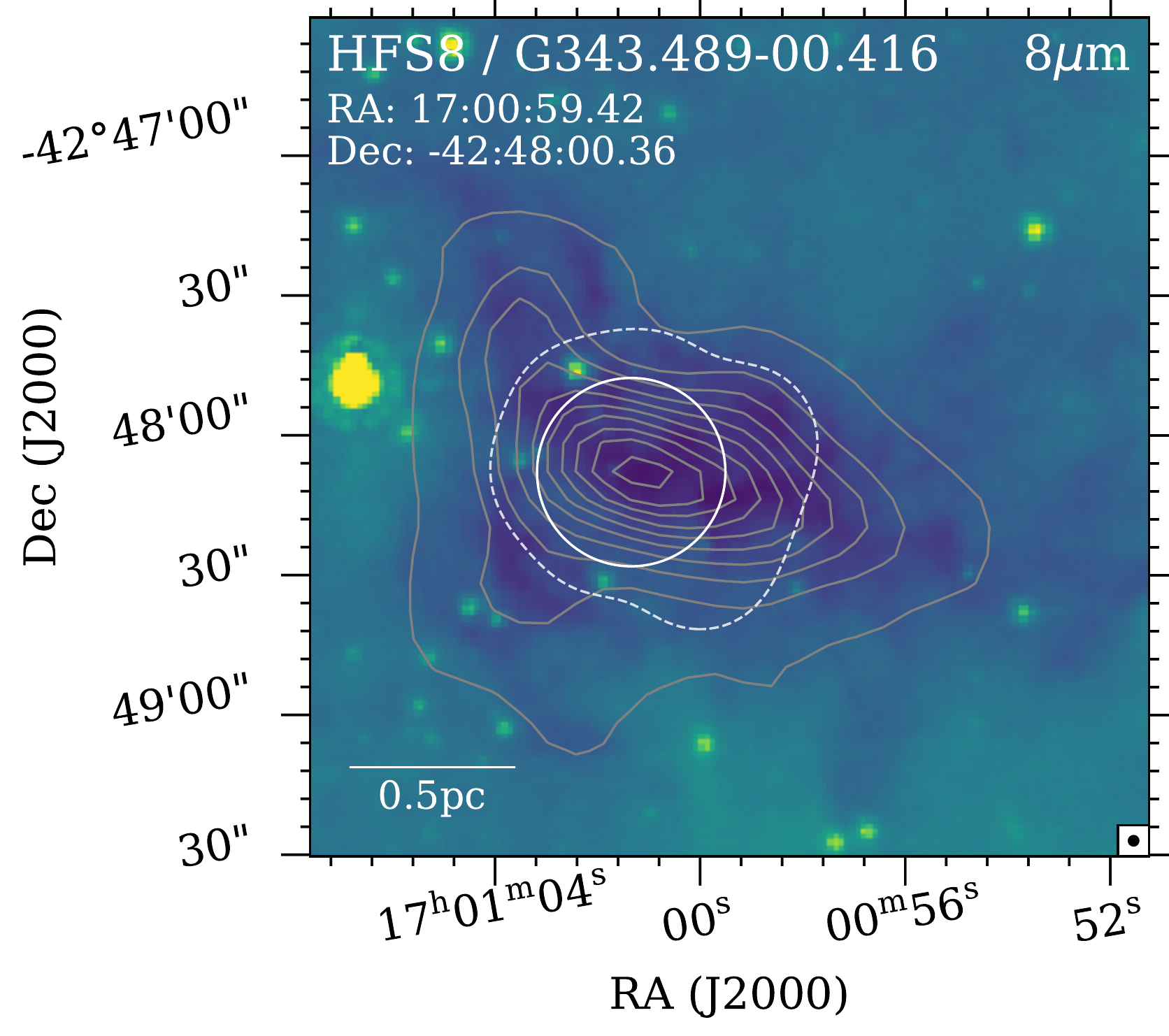}
\includegraphics[width=2.0 in]{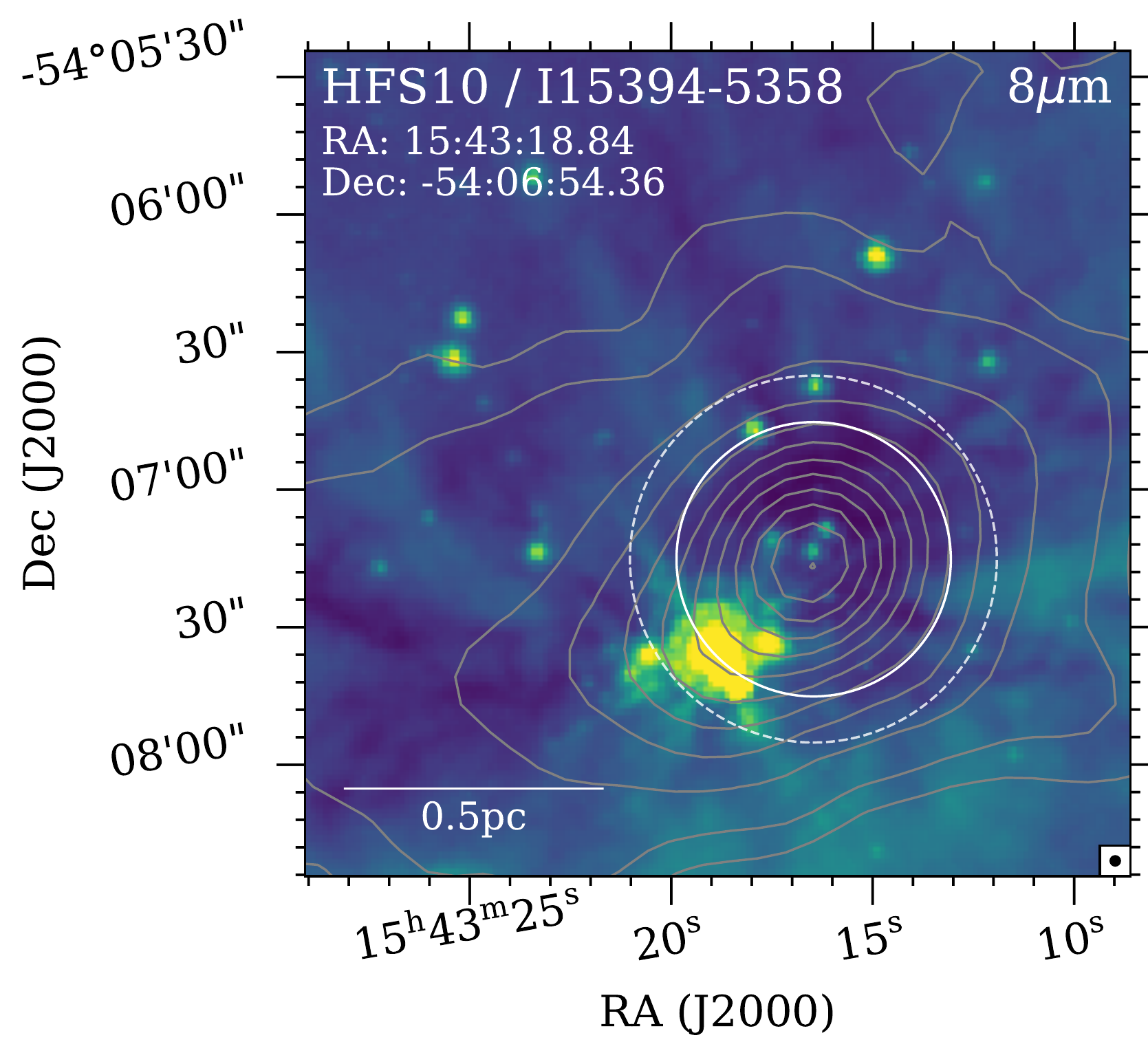}
\includegraphics[width=2.0 in]{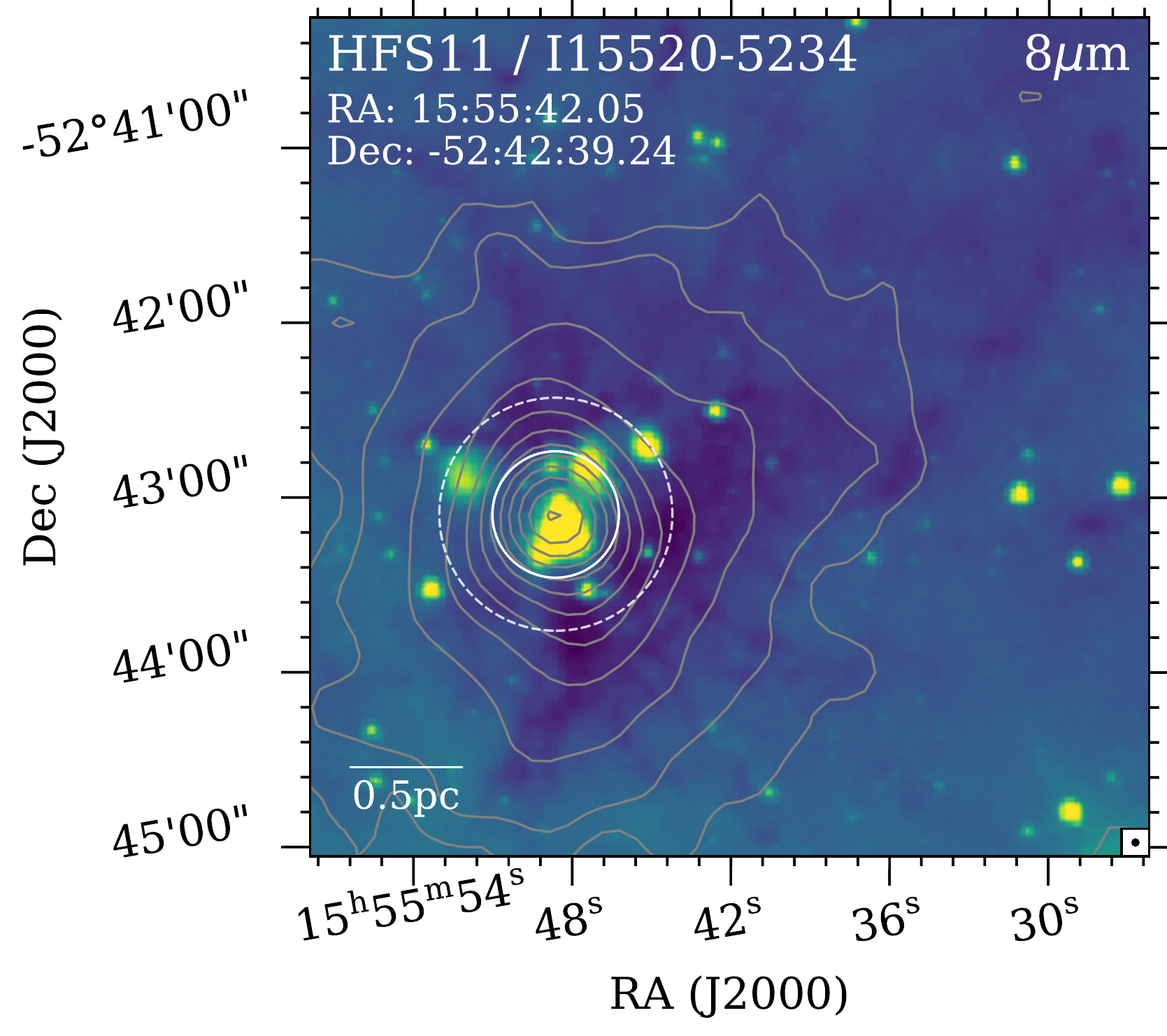}
\includegraphics[width=2.0 in]{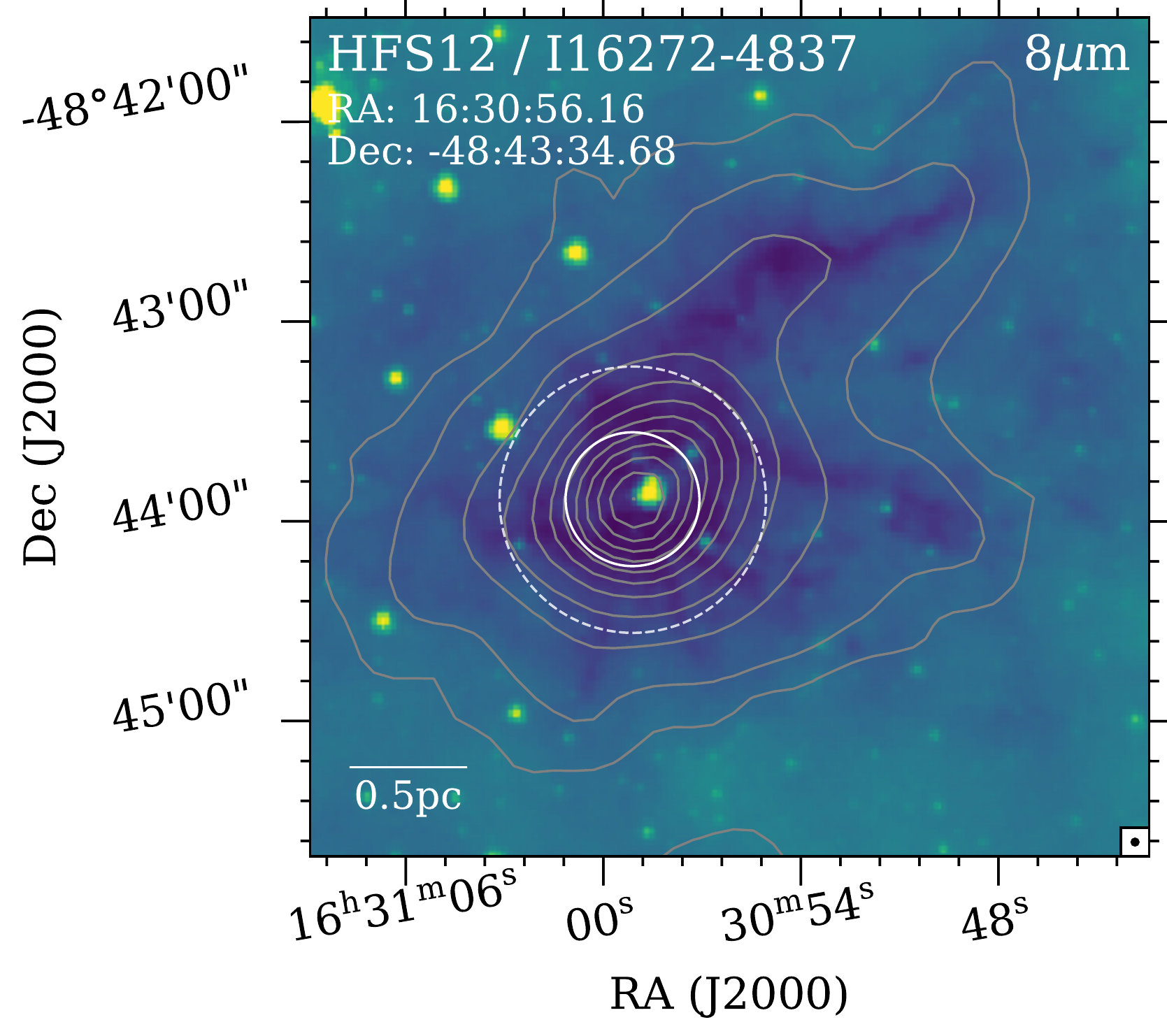}
\includegraphics[width=2.0 in]{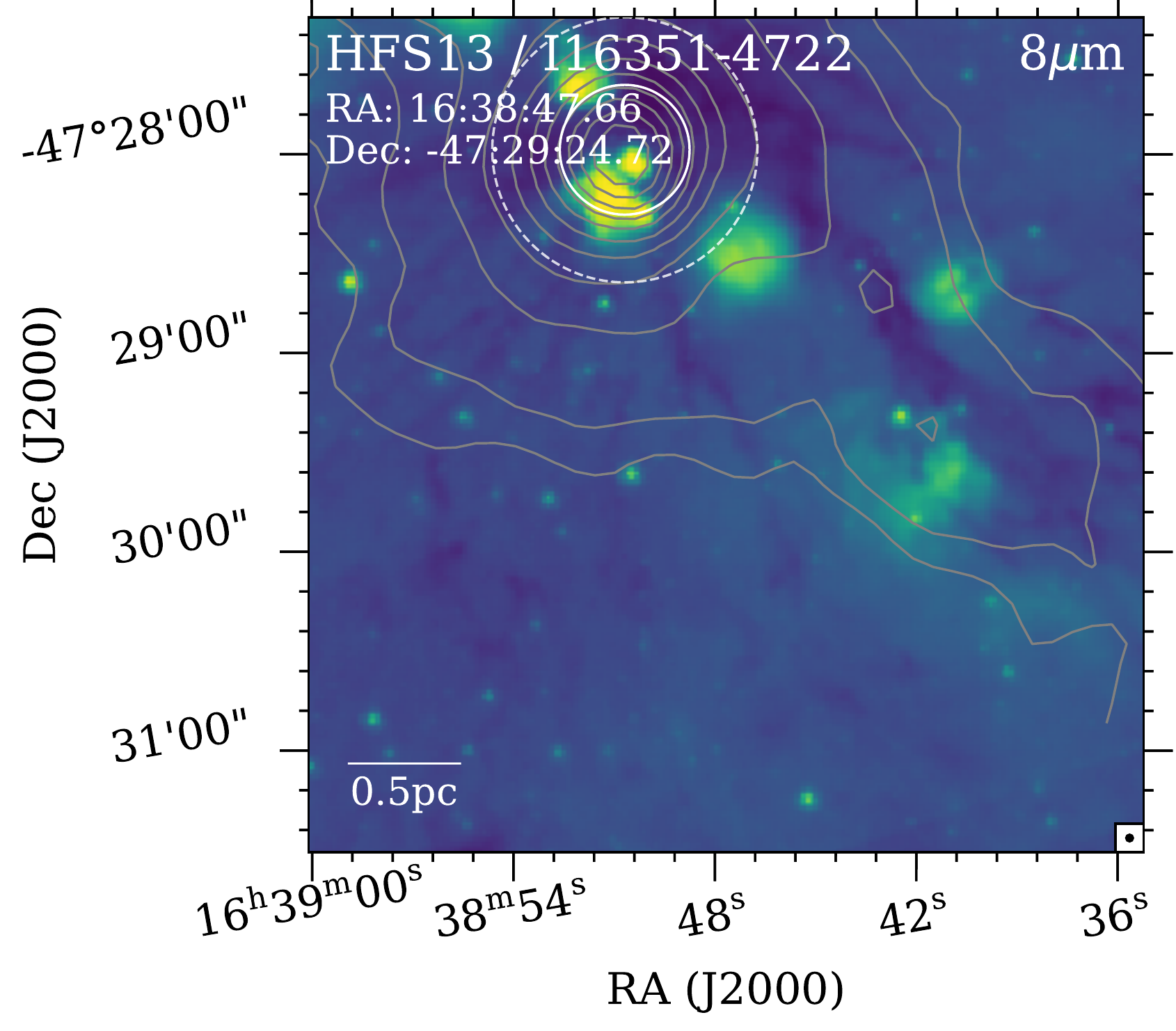}
\includegraphics[width=2.0 in]{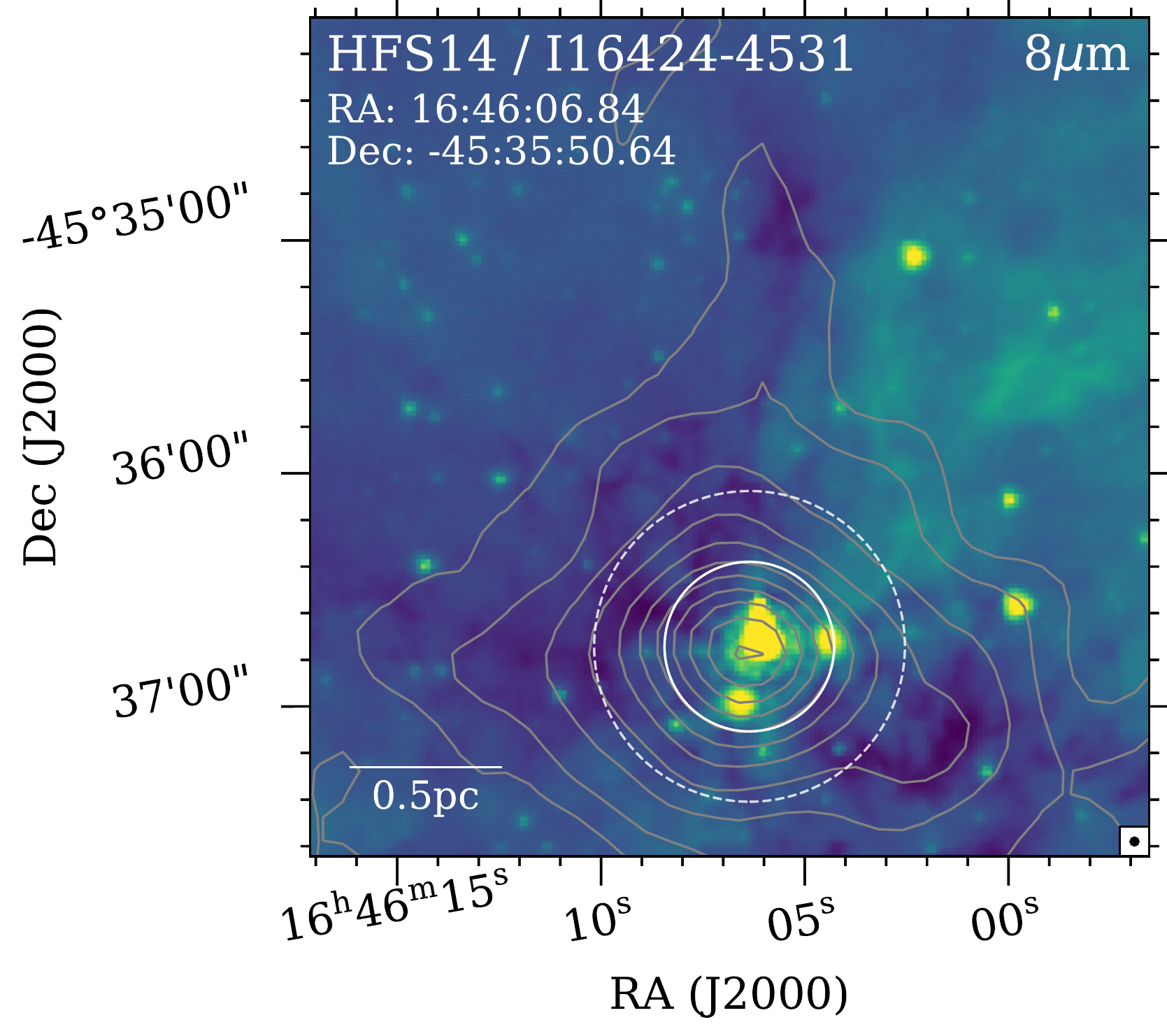}
\includegraphics[width=2.0 in]{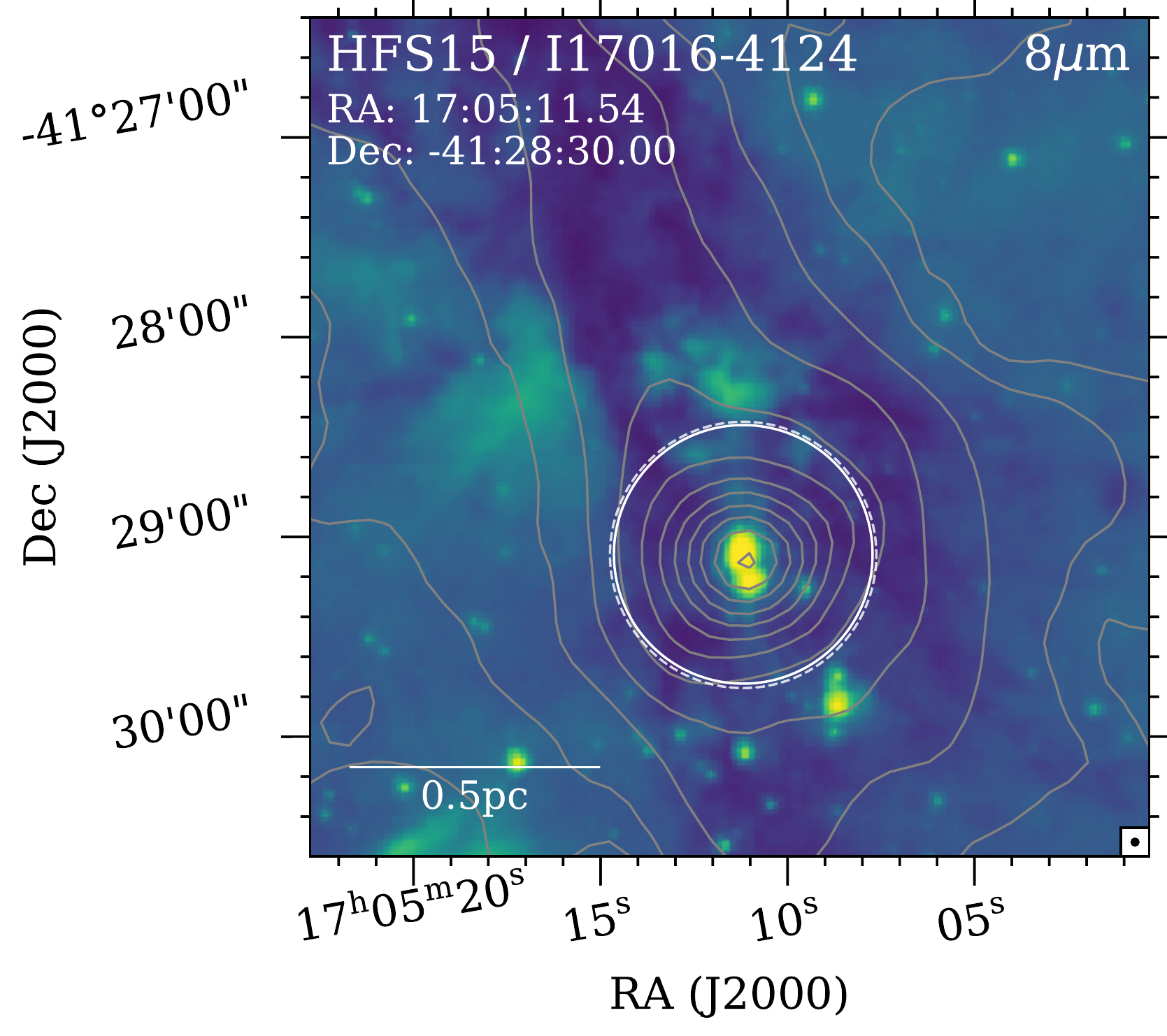}
\includegraphics[width=2.0 in]{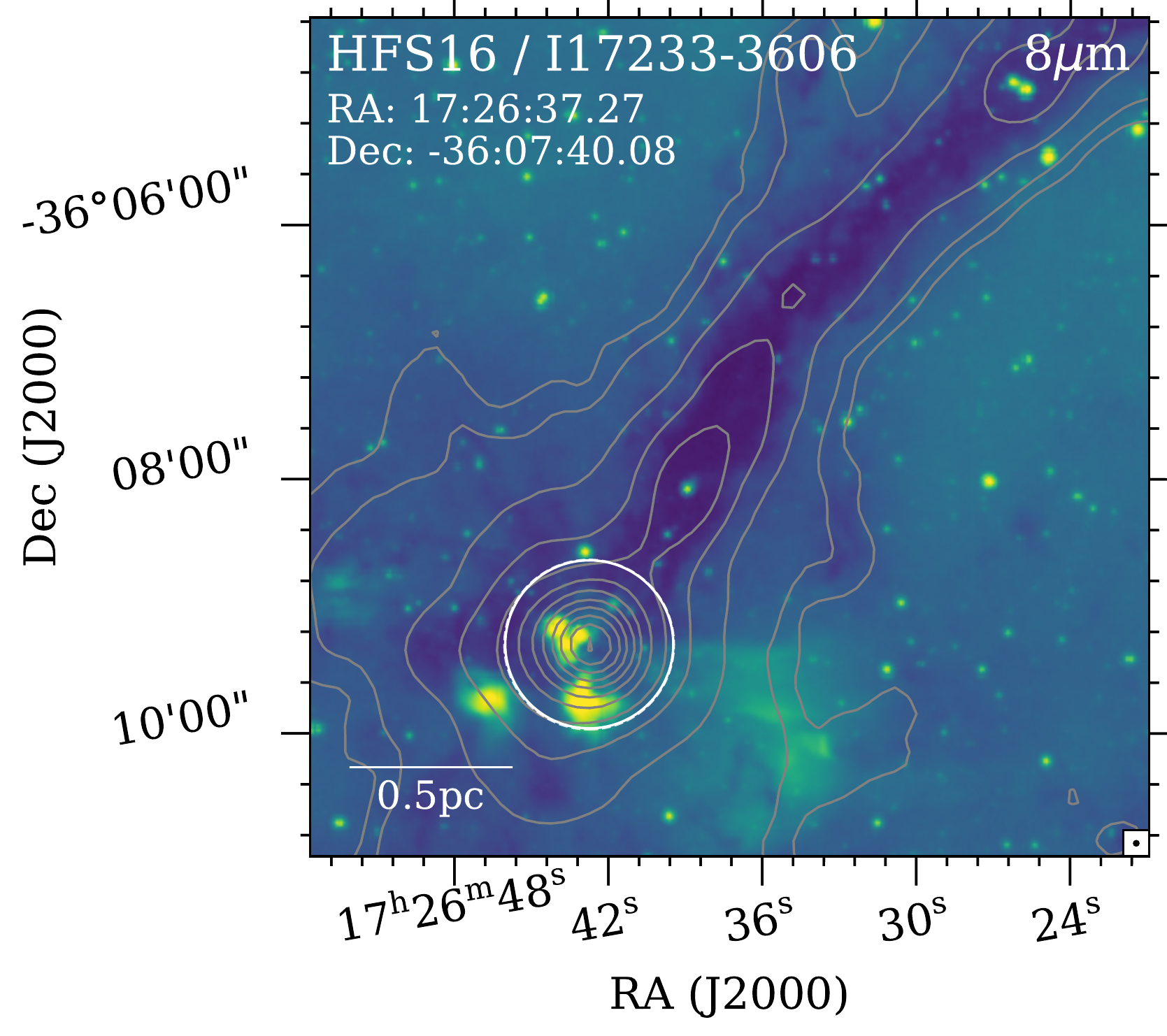}
\includegraphics[width=2.0 in]{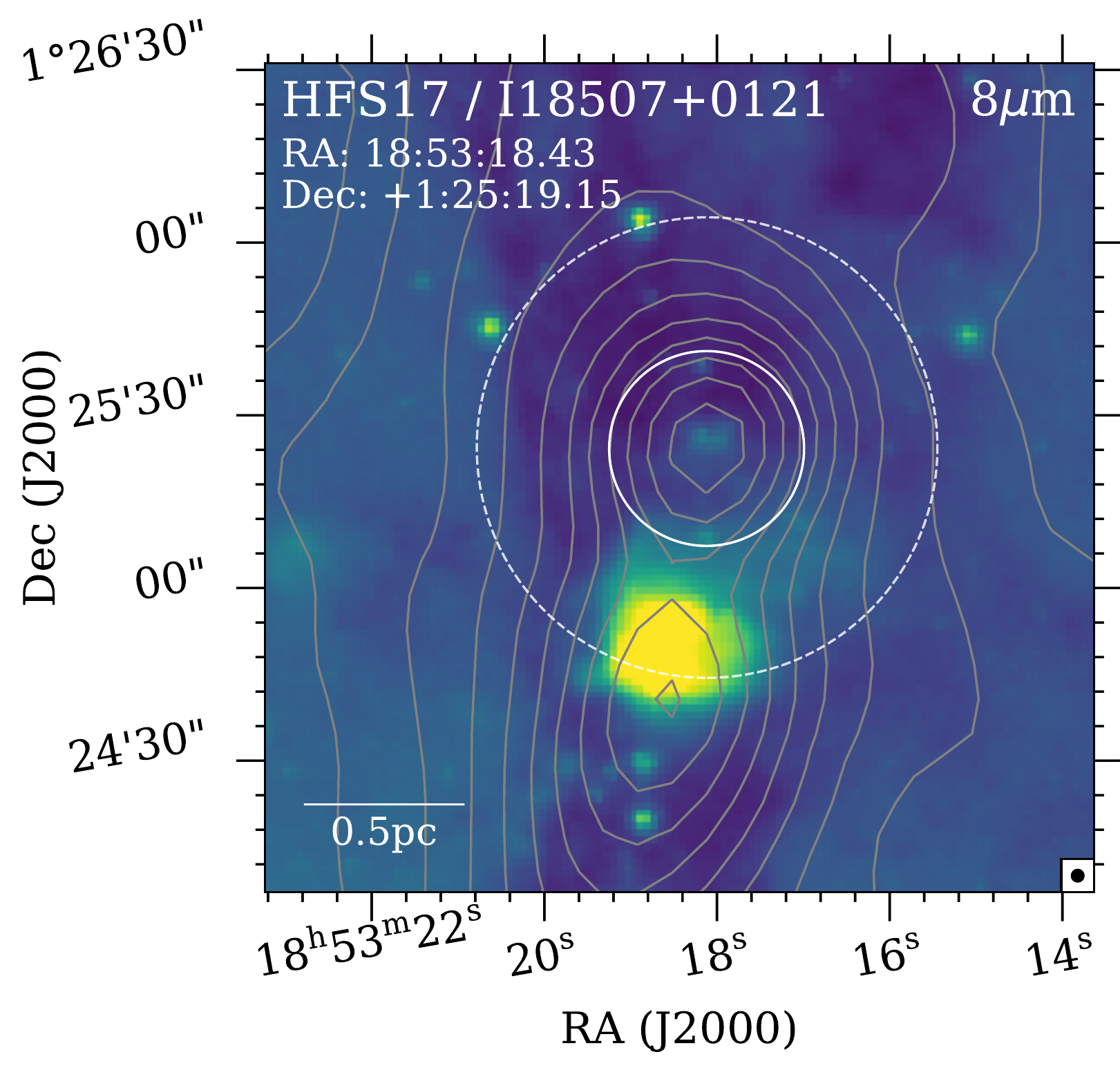}
\caption{Same as Fig.\,\ref{fig:overview_hfs} but for the remaining
15 HFSs. 
}
\label{fig:overview_hfs_app}
\end{figure*}


\vspace{5mm}
\noindent
Author affiliations:\\

\noindent 
$^{1}$School of physics and astronomy, Yunnan University, Kunming, 650091, PR China \\
$^{2}$Indian Institute of Space Science and Technology, Thiruvananthapuram 695 547, Kerala, India\\
$^{3}$Shanghai Astronomical Observatory, Chinese Academy of Sciences, 80 Nandan Road, Shanghai 200030, Peoples Republic of China \\
$^{4}$Key Laboratory for Research in Galaxies and Cosmology, Shanghai Astronomical Observatory, Chinese Academy of Sciences, 80 Nandan Road, Shanghai 200030, Peoples Republic of China \\
$^{5}$National Astronomical Observatory of Japan, National Institutes of Natural Sciences, 2-21-1 Osawa, Mitaka, Tokyo 181-8588, Japan \\
$^{6}$Department of Astronomical Science, The Graduate University for Advanced Studies, SOKENDAI, 2-21-1 Osawa, Mitaka, Tokyo 181-8588, Japan \\
$^{7}$Yunnan Observatories, Chinese Academy of Sciences, 396 Yangfangwang, Guandu District, Kunming, 650216, China\\
$^{8}$Chinese Academy of Sciences South America Center for Astronomy, National Astronomical Observatories, CAS, Beijing 100101, China\\
$^{9}$Departamento de Astronom\'ia, Universidad de Chile, Casilla 36-D, Santiago, Chile\\
$^{10}$Jet Propulsion Laboratory, California Institute of Technology, 4800 Oak Grove Drive, Pasadena, CA 91109, USA\\
$^{11}$Department of Astronomy, Graduate School of Science, The University of Tokyo, 7-3-1 Hongo, Bunkyo-ku, Tokyo 113-0033, Japan \\
$^{12}$Max Planck Institute for Astronomy, Königstuhl 17, D-69117 Heidelberg, Germany \\
$^{13}$Departamento de Astronom\'ia, Universidad de Concepci\'on, Av. Esteban Iturra s/n, Distrito Universitario, 160-C, Chile \\
$^{14}$Max-Planck-Institute for Astronomy, K\"{o}nigstuhl 17, 69117 Heidelberg, Germany \\
$^{15}$Kavli Institute for Astronomy and Astrophysics, Peking University, 5 Yiheyuan Road, Haidian District, Beijing 100871, People's Republic of China\\
$^{16}$Department of Astronomy, Peking University, 100871, Beijing, People's Republic of China\\
$^{17}$Indian Institute of Astrophysics, Koramangala II Block, Bangalore 560 034, India\\
$^{18}$Satyendra Nath Bose National Centre for Basic Sciences, Block-JD, Sector-III, Salt Lake, Kolkata-700 106 \\
$^{19}$E\"{o}tv\"{o}s Lor\'{a}nd University, Department of Astronomy, P\'{a}zm\'{a}ny P\'{e}ter s\'{e}t\'{a}ny 1/A, H-1117, Budapest, Hungary\\
$^{20}$Physical Research Laboratory, Navrangpura, Ahmedabad—380 009, India \\
$^{21}$National Astronomical Observatories, Chinese Academy of Sciences, Beijing 100101, China  \\
$^{22}$Korea Astronomy and Space Science Institute, 776 Daedeokdaero, Yuseong-gu, Daejeon 34055, Republic of Korea\\
$^{23}$University of Science and Technology, Korea (UST), 217 Gajeong-ro, Yuseong-gu, Daejeon 34113, Republic of Korea\\
$^{24}$School of Physics and Astronomy, Sun Yat-sen University, 2 Daxue Road, Zhuhai, Guangdong, 519082, People's Republic of China\\
$^{25}$SOFIA Science Centre, USRA, NASA Ames Research Centre, MS-12, N232, Moffett Field, CA 94035, USA \\

\label{lastpage}
\end{document}